\documentclass{jfm}
\pdfoutput=1

\usepackage{graphicx}
\usepackage{epstopdf, epsfig,grffile}
\usepackage{bm}
\usepackage{amsmath}
\usepackage{url}
\usepackage{color}
\usepackage{comment}
\usepackage{natbib}

 \usepackage{algorithmicx}
  \usepackage[ruled]{algorithm}
  \usepackage{algpseudocode}

\newcommand{\ba}{\boldsymbol{a}}
\newcommand{\bi}{{\boldsymbol{i}}}
\newcommand{\bj}{{\boldsymbol{j}}}

\newcommand{\bsigma}{{\boldsymbol{\sigma}}}

\newcommand{\bv}{{\boldsymbol{v}}}

\newcommand{\bX}{{\boldsymbol{X}}}
\newcommand{\bx}{{\boldsymbol{x}}}

\newcommand{\bz}{{\boldsymbol{z}}}
\newcommand{\bZ}{{\boldsymbol{\gamma}}}

\newcommand{\mA}{{\mathcal{A}}}
\newcommand{\mB}{{\mathcal{B}}}
\newcommand{\mD}{{\mathcal{D}}}
\newcommand{\mS}{{\mathcal{S}}}

\newcommand{\tf}{{t_\mathrm{f}}}

\newcommand{\id}{\text{\bf I}}

\newcommand{\add}[1]{\textcolor{black}{#1}}
\newcommand{\del}[1]{{ }}

\shortauthor{S. Thalabard, J. Bec}

\title[Turbulence of generalised flows in two dimensions]{Turbulence of generalised flows\\ in two dimensions}

\author{Simon Thalabard \aff{1}\corresp{\email{simon.thalabard@ens-lyon.org}} \and J\'{e}r\'{e}mie Bec\aff{2}}

\affiliation{
\aff{1}
Instituto Nacional de Matem\'atica Pura e Aplicada, IMPA, 22460-320 Rio de Janeiro, Brazil
\aff{2} MINES ParisTech, PSL Research University, CNRS, CEMEF, Sophia-Antipolis, France}

\begin{document}

\maketitle

\begin{abstract}
This paper discusses the generalised least-action principle and the associated concept of generalised flow introduced by 
Brenier (\emph{J.\ Am.\ Math.\ Soc.}\/ 1989),  from the perspective of turbulence modelling. In essence, Brenier's generalised least-action principle  extends to a probabilistic setting Arnold's geometric interpretation of ideal fluid mechanics, whereby strong solutions to the Euler equations are deduced from minimising an action over Lagrangian maps. While Arnold's framework relies on the deterministic concept of Lagrangian flow,  Brenier's least-action principle describes solutions to the Euler equations  in terms of non-deterministic \emph{generalised flows}, namely probability measures over sets of Lagrangian trajectories.

 In concept, generalised flows  seem naturally fit to describe turbulent Lagrangian trajectories in terms of  stochastic processes, an approach that originates from Richardson's seminal work on turbulent dispersion.
\add{Still, Brenier's generalised least-action principle has so far hardly found any practical application in the realm of fluid mechanics, let alone for turbulence modelling. } 
\add{ Naturally, one would ultimately desire  to use generalised flows  to address scenarios when  the Lagrangian flow breaks down, due to Lagrangian trajectories being \emph{spontaneously stochastic}. 
A first step in that direction is to provide a hydro-dynamical perspective on  Brenier's principle in cases where  such exotic behaviours are ruled out,  and where Lagrangian  trajectories are univocally defined at a microscopic level.
Specifically, we wish to assess the  potential skills of the generalised least-action principle at  coarse-graining the  Lagrangian motion  of deterministic fluid particles, and reconstructing the Eulerian space-time dynamics of the associated velocity fields.}

\add{In practice,  we rely on  a statistical-mechanics interpretation of the concept of generalised flows, whereby the latter becomes akin to statistical ensembles of suitably defined ``permutation flows''. We then employ Monte-Carlo techniques to numerically solve the generalised least-action principle, and analyse the Eulerian and Lagrangian statistical features of the associated generalised flows. For simplicity, we restrict ourselves to two dimensions of space and consider situations of increasing complexities, ranging from solid rotation and cellular flows to freely decaying two-dimensional turbulence.}

\add{
Our analysis highlights  a major caveat of  the generalised variational principle, that produces non-physical dynamics when used over long time lags, \emph{e.g.}, longer than  a well-defined \emph{hydrodynamic turnover time}. We argue that this limitation is not specifically inherent to Brenier's formulation, but rather to the variational framework being formulated as a two-end boundary-value problem.  When appropriately used over sufficiently short times,  generalised flows are shown to be relevant, as they  may even reproduce irreversible Eulerian behaviors.  This suggests that if carefully used, generalised variational formulations could   provide new tools to coarse-grain genuine multi-scale  hydrodynamics.}

  
  
\end{abstract}

\graphicspath{}


\section{Introduction}
In the Lagrangian framework, incompressible ideal fluids can be pictured as freely evolving space-filling assemblies of infinitely-many classical ``fluid particles'', whose internal energies are neglected, and whose collective dynamics is therefore conservative. It may hence come as no surprise that the Euler equations can be formally deduced from a Lagrangian least-action principle, whereby the action is the time integral of the total kinetic energy,  while incompressibility  appears as a non-holonomic constraint for the Lagrangian map $(\boldsymbol{a},t)\mapsto\boldsymbol{X}(\boldsymbol{a},t)$, which is assumed to be a smooth function of both time and the Lagrangian coordinates $\boldsymbol{a} $.

The deterministic Lagrangian least-action principle was made rigorous by \cite{arnold1966geometrie}\,---\,see also \cite{khesin2005topological,arnold2013mathematical,farazmand2018variational}. It is at the root of the subsequent classical literature on Hamiltonian formulation of ideal fluid dynamics, which has found widespread ramifications in the context of geophysical and plasma modelling\,---\,we refer the reader to the pedagogical reviews by \cite{shepherd1990symmetries}, \cite{morrison1998hamiltonian}, \cite{salmon1988hamiltonian}, and references therein. To briefly emphasise the power of least-action principles, let us here just point out that Hamiltonian methods not only provide a practical guideline that can be used to control approximations in asymptotic models \citep{holm1983noncanonical,salmon1983practical,salmon1985new,morrison2005hamiltonian} and develop faithful numerical methods \citep{marsden2001discrete,zeitlin2004self,kraus2016variational}, but also lay the basis for systematic stability analysis \citep{holm1985nonlinear,abarbanel1986nonlinear} and formal statistical mechanics computations \citep{robert1991statistical,miller1992statistical,bouchet2010invariant,bouchet2012statistical}.  In spite of its relevance when formally putting viscosity to zero, Arnold's Lagrangian formulation is not necessarily appropriate in the context of turbulence, that is to describe solutions to the Navier--Stokes equations in the limit of vanishing viscosity.\\

From a mathematical viewpoint, Arnold's least-action principle yields spatially smooth solutions to the Euler equations.  Yet, it is well known that the notion of solution to the Euler equations needs to be carefully weakened in order to craft velocity fields that reproduce hallmark turbulent features while remaining physical \citep[see, e.g.,][]{duchon2000inertial,eyink2002local}.  To that end, the notion of distributional solution with prescribed H\"older spatial regularity has been particularly scrutinised in relation to a conjecture made by \cite{onsager1949statistical}\,---\,see also \cite{eyink2006onsager}. It has been recently established \citep{buckmaster2015onsager,buckmaster2016dissipative,isett2016proof} that dissipative distributional solutions to the \add{3D} Euler equations do exist with a spatial H\"older exponent slightly below $1/3$. This indicates that an inviscid dynamics could indeed be relevant to reproduce experimental measurements, including the persistence of a finite dissipation in the limit of infinite Reynolds number \citep[the so-called ``dissipative anomaly'', see][]{sreenivasan1984scaling}, the 4/5 law \citep[see, e.g.,][]{antonia2006approach}, and possibly deviations to Kolmogorov 1941 scaling \citep[see, e.g.,][]{saw2018universality}. It cannot be ruled out that  constructing  physical \add{turbulent} solutions to the Euler equations  requires considering even weaker notions of solutions, examples of which include measure-valued solutions of \cite{diperna1987oscillations}\,---\,see also \cite{brenier2011weak}\,---\,where the local velocity is not uniquely defined but rather prescribed by a local probability distribution, namely a Young measure.

At a physical level, a crucial consequence of the velocity being probabilistic is that fluid-element trajectories become themselves probabilistic: The concept of Lagrangian flow breaks down. A similar consequence is conjectured to also hold true for the rough velocity fields imagined by Onsager: This relates to the phenomenon of Lagrangian ``spontaneous stochasticity'' that was first formalized in the context of Kraichnan's model for turbulent advection by \cite{gawedzki2001turbulent}\,---\,see also \cite{falkovich2001particles}.  Clearly, a Lagrangian least-action principle built from the deterministic notion of Lagrangian map cannot reproduce such an expected intrinsic stochasticity of fluid trajectories.  To that end, the generalised Lagrangian least-action principle formulated by \cite{brenier1989least} could however be of particular relevance.  Brenier's formulation is essentially a probabilistic generalisation of Arnold's principle.  Roughly said, it consists in minimising a suitably-defined average kinetic energy among a very wide class of stochastic processes\,---\,the so-called ``generalised Lagrangian flows''\,---\,rather than among the sole deterministic class of smooth diffeomorphisms.  \add{From a mathematical perspective,} this generalisation guarantees the existence of minimisers under rather weak assumptions. The minimising \emph{generalised flow} is \add{no longer a map} but rather defined as a probability distribution on the space of Lagrangian paths. 
The solution is obtained by prescribing boundary conditions in time, which amounts to specifying the Lagrangian transition probabilities from the initial to the final time. 
\add{Let us note that this formulation of the problem remains compatible with deterministic Lagrangian motion. In that case, the minimising flow is degenerate and formally described by a Dirac measure, and  it was shown  by \cite{brenier1989least} that Arnold's solutions are indeed retrieved by the generalised variational principle. This essentially means that for genuinely probabilistic generalised flows to appear, Arnold's principle has to break down.
Let us also remark, that Brenier's framework is rooted in optimal transport and control theories \citep[see for instance][]{villani2008optimal}: Brenier's formulation ressembles a Kantorovich optimisation while Arnold's formulation is essentially ``Monge-like''.}

\add{Brenier's generalised flows appear to be very versatile: They encompass the smooth solutions to the Euler equations produced by Arnold's principle on the one hand, and on the other hand they turn out to provide a framework for explicit constructions of weak dissipative Euler solutions, with possible relevance to turbulence  \citep{shnirelman2000weak}. 
Besides, Brenier's least-action principle is a practical principle: It produces  generalised flows that formally connect to the Euler equations, and which can be explicitly constructed through various numerical methods used in multi-marginal optimal transport \citep{benamou2017generalized,merigot2016minimal,gallouet2016lagrangian}. 
Beyond the mathematical framework developed in \citep{brenier1989least,brenier1999minimal,brenier2008generalized}, the question remains as to whether the generalised flows  solving Brenier's least-action principle have any physical relevance for hydro-dynamics, let alone for turbulence modelling. One could formulate the question in the following manner: When the generalised minimisers do not coincide with classical solutions and are then likely non-deterministic,  can their statistical  features  be interpreted in terms of a transition towards a turbulent state, hereby making the principle compatible with the conjectured turbulent breakdown of Lagrangian flows?
The purpose of the present paper is to partially address this question based on a physical understanding of Brenier's generalised least action principle. }

\add{
A thorough assessement on the physical relevance of Brenier's least action principle  is out of our scope but should necessarily be nuanced.
The intrinsic statistical features of the minimising flows depend on the input data: In Breniers' formulation it is specified in terms of  a two-end coupling, namely in terms of Lagrangian transition probabilities between initial and final times.
If the input data is generated from Euler dynamics, one should  rightfully wish that solving the generalised least-action principle yields back the full solution. This is indeed the case for short times, as proven by \cite{brenier1989least}.
Yet, having the mind set on data reconstruction, one woud rather consider a physical scenario, where the input data would be generated by viscous dynamics (Navier--Stokes) and would necessarily be \emph{coarse-grained} in space.
In those cases, one could still solve for the minimising generalised flows, but it is then unclear how relevant or even how physical those would be. Considering some sufficiently simple yet non-trivial data  however yields some fruitful physical insights on those matters.  }

\add{Our intention is to assess the potential skills of the generalised least-action principle at coarse-graining the Lagrangian motion of fluid particles, and reconstructing therefrom the Eulerian space-time dynamics of the associated velocity fields.  For simplicity, we only consider cases, where exotic behaviours such as ``explosive separation'' of trajectories are ruled out: The Lagrangian trajectories are univocally defined at a microscopic level. While spatial coarse-graining still calls for a probabilistic description of trajectories, we may \emph{a priori} expect the minimising flows to be only midly non-degenerate, at least if the Lagrangian trajectories are non-chaotic.  For this reason, it is enough to restrict our analysis to two dimensions of space, where we yet consider situations of increasing complexities, ranging from solid rotation and cellular flows to freely decaying two-dimensional turbulence.}

\add{In practice, we rely on a statistical-mechanics interpretation of the concept of generalised flows, whereby the latter become akin to statistical ensembles of suitably defined``permutation flows''. We then employ Monte-Carlo techniques to numerically solve the generalised least-action principle, and analyse the Eulerian and Lagrangian statistical features of the associated generalised flows.  Our analysis highlights a major caveat of the generalised variational principle, that produces non-physical dynamics when solved over long time intervals, \emph{e.g.}, longer than a well-defined hydrodynamic turnover time. We argue that this limitation is not specifically inherent to Brenier's formulation, but rather to the variational framework being formulated as a two-end boundary-value problem: When solved over the same time lags, Arnold's least action principle also produces incorrect Euler solutions.  When appropriately used over sufficiently short times, the generalised least-action principle is shown to be relevant, for all the examples considered. Besides, our third example shows that it also succeeds at reproducing irreversible Eulerian behaviors, such as the enstrophy and energy decays of 2D decaying turbulence. This suggests that if carefully used, generalised variational formulations could provide new tools to coarse-grain genuine multi-scale hydrodynamics.}

The paper is organised as follows. In \S\ref{sec:variational} we formulate Arnold's and Brenier's least-action principles in a fluid-mechanical language. In \S\ref{sec:statphys} we provide a discrete version of generalised flows that allows for numerical constructions based on Monte-Carlo techniques. In \S\ref{sec:Beltrami} we use this method with, as input data, a coarse-grained Lagrangian map associated to Beltrami (cellular) flows.  By investigating the time variations of kinetic energy, the scaling properties of the velocity field and Lagrangian trajectories, we argue that generalised flows display non-physical features when the time lag between the initial and final times is too large. We then use such considerations to investigate the case of two-dimensional decaying turbulence in \S\ref{sec:decay}. We observe that the generalised least-action principle allows to reconstruct the coarse-grained velocity when the time lag is chosen of the order or smaller than the large-eddy turnover time. Finally, we draw concluding remarks in \S\ref{sec:conclusion}.


\section{Lagrangian variational formulations of inviscid fluid dynamics}
\label{sec:variational}

\subsection{Arnold's least-action principle}
In order to set the framework for Brenier's formulation, let us first formally restate Arnold's variational formulation of ideal fluid dynamics. Beyond the virtue of setting up some notations, we will then already be in a position to comment on the existence of a critical time-lag $t_\star$, over which Arnold's \emph{least-action} principle becomes a weaker principle of \emph{stationary action}: This unwanted feature will have a counterpart in Brenier's formulation.
To focus on general physical ideas, our exposition is intentionally phenomenological, and we refer the mathematically inclined reader to the book of \cite{khesin2005topological} and the review of \cite{ebin1970groups}.\\

Arnold's formulation is rooted on the deterministic Lagrangian viewpoint. The fluid is then nothing but the collection of individual fluid particles, which are labelled by some Lagrangian set of coordinates $\ba$, and whose trajectories determine its dynamics. The formalisation uses the concept of Lagrangian flow map $(\ba,t) \mapsto\bX(\ba,t)$, which associates to each fluid particle its position at time $t$ within the physical domain ${\mD}$.  The projections $\ba \mapsto \bX(\ba,t)$ define the Lagrangian maps at time $t$, and the projections $t\mapsto \bX(\ba,t)$ define the trajectories of particles with Lagrangian label $\ba$.  The Lagrangian flows are assumed to be smooth functions of both time and the Lagrangian coordinates.  It is convenient to identify Lagrangian labels with the initial positions of fluid elements, so that $\bX(\ba,0) = \ba$.  Hence, in the sequel, the initial map will always be taken as identity.

Lagrangian flows describe incompressible fluids when they preserve volumes, that is when the Jacobian determinant of the Lagrangian map satisfies $|\partial \bX/\partial \ba|\equiv 1$. The Eulerian velocity field is then obtained from the particle velocities through $\bv(\bX(\ba,t),t) =\partial_t \bX(\ba,t)$. The total fluid kinetic energy therefore reads
\begin{equation}
  \mathcal{E}(t) := \int_\mD \mathrm{d}\ba\, \frac{1}{2}
\|\partial_t\bX(\ba,t)\|^2.
\end{equation}
Arnold's Variational Principle can now be stated as
\begin{equation}
  \tag{AVP}
  \begin{split}
    \mathcal{A}_{0, \tf} [\boldsymbol{X}(\cdot)] :=
    \int_{0}^{\tf}\mathrm{d}t \,\mathcal{E}(t)  \longrightarrow \text{inf} \qquad \text{subject to}
    &\\
     \bX(\cdot,0) = \id,\ \  \bX(\cdot,\tf)=\bX_{\rm f}(\cdot), \mbox{ and } \left|\frac{\partial \bX}{\partial\ba}\right|\equiv 1.&
  \end{split}
  \label{eq:AVP}
\end{equation}
The action functional $\mathcal{A}_{0,\tf}$ appears as a natural counterpart to the actions used in textbooks for the formulation of classical finite-dimensional mechanics \citep[see, e.g.,][]{jose2000classical}.  It is obvious but worth emphasising, that the formulation (\ref{eq:AVP}) defines a \emph{boundary-value} optimisation problem, as both ``end-point'' maps $\bX(\cdot,0)=\id$ and $\bX_{\rm f}(\cdot)$ at final time $\tf$ are prescribed.  A formal way of solving (\ref{eq:AVP}) is to use a Lagrange multiplier $P(\ba,t)$ to enforce the incompressibility constraint, and compute the functional derivatives with respect to the flow $\bX$ of the augmented action $A_{0,\tf} + \int_{0}^{\tf} \,\mathrm{d}t \int_\mD \mathrm{d}\ba \, P(\ba,t) \,(|\partial \bX/\partial \ba|-1)$. The optimal solutions necessarily solve the Euler-Lagrange equations. Requiring the action to be stationary ($\delta\mathcal{A}_{0,\tf} =0$), a string of formal calculations leads to
\begin{equation}
  \partial_{tt} \bX(\ba,t)=- \left[{\partial \bX}/{\partial\ba}\right]^{-1}\nabla_{\ba} P, \quad
  \mbox{ with }  \left|{\partial \bX}/{\partial\ba}\right|\equiv 1,
  \label{eq:euler_lag}
\end{equation}
which is exactly the Lagrangian formulation of the inviscid dynamics of an incompressible fluid. Note that Eq.~(\ref{eq:euler_lag}) is second order in time, which means that we need two boundary conditions provided by the initial and final maps in Arnold's principle (\ref{eq:AVP}).

\subsection{Boundary-value formulation and turnover time}
\label{ssec:arnold-solid}
The boundary-value formulation of Arnold's principle (\ref{eq:AVP}) is not usual in fluid mechanics. One rather considers the initial-value problem, which involves the initial fluid velocity $\partial_t \bX(\ba,0) = \bv_0(\ba)$ in addition to the initial map. However, as we will now see, the connection between these two formulations is not univocal.

\begin{figure}
  \begin{minipage}{0.25\textwidth}
    \begin{center}
      \includegraphics[width=\textwidth]{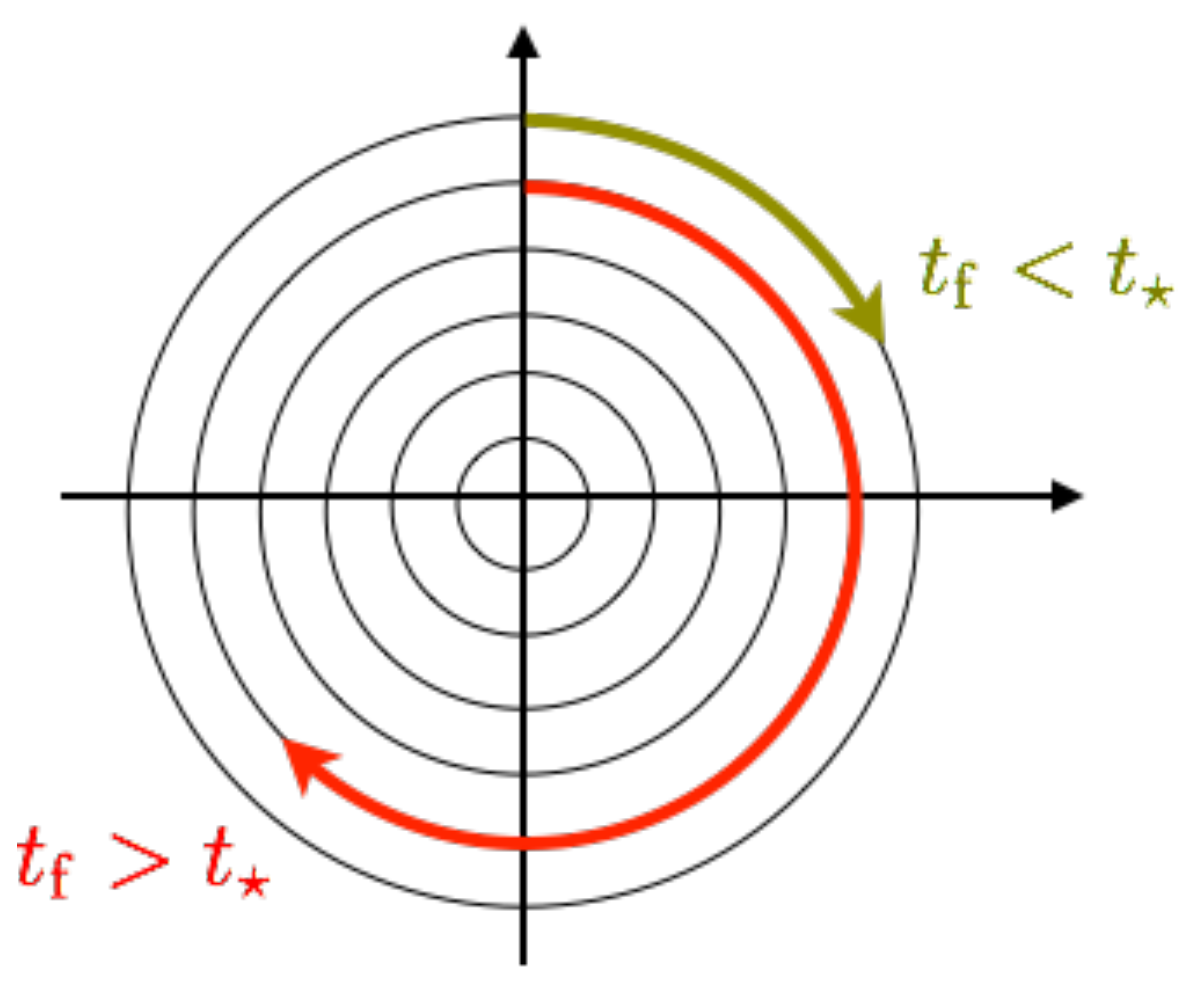}\\[20pt]
      \strut
    \end{center}
  \end{minipage}
  \qquad
  \begin{minipage}{0.6\textwidth}
    \includegraphics[width=\textwidth]{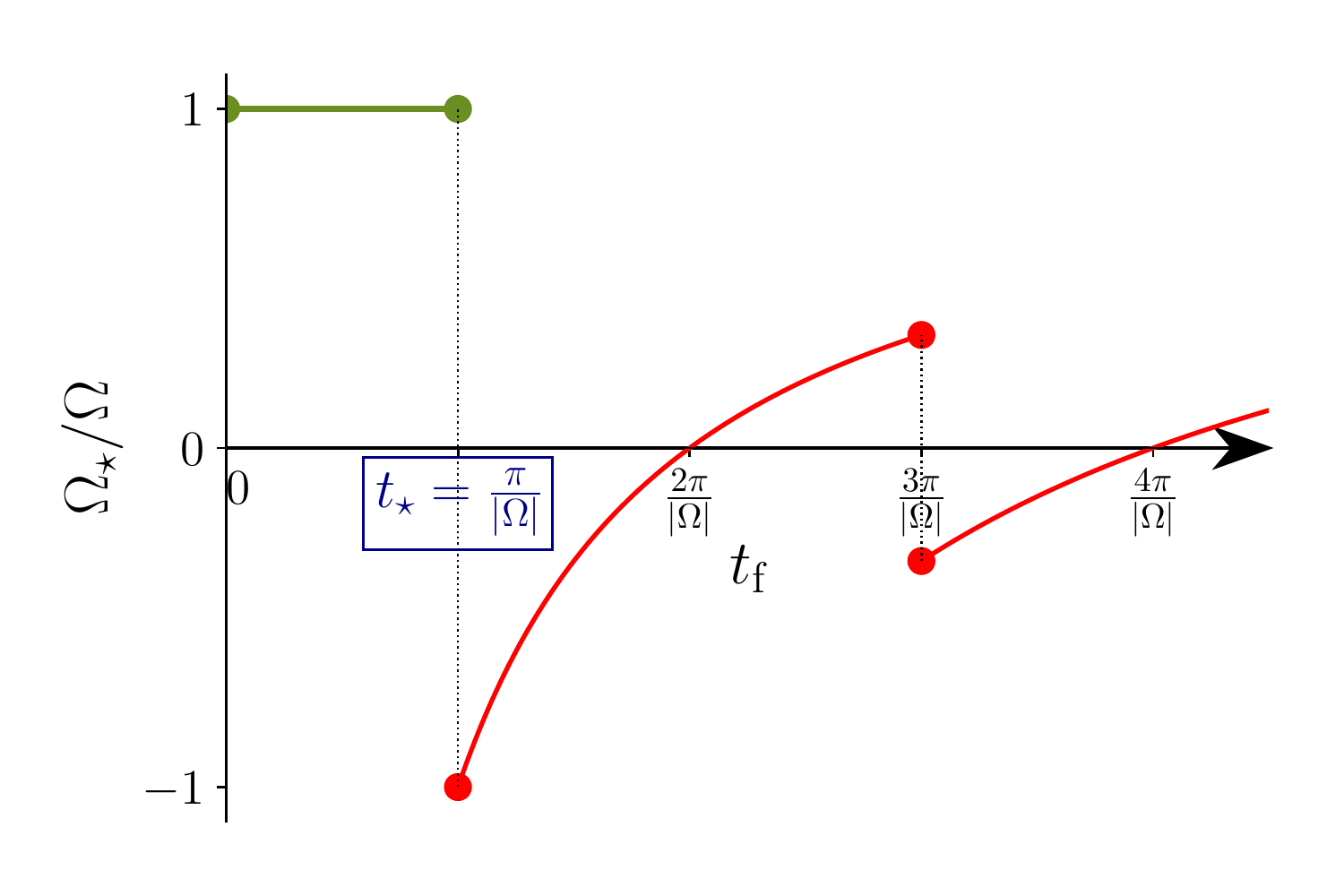}
  \end{minipage}
  \caption{\add{A solid-rotating flow with pulse $\Omega$ only solves the variational principle (\ref{eq:AVP}) up to final times $\tf<t_\star(\Omega)$ (in green),  when the final map has rotated less than half a turn (left illustration). For $\tf>t_\star$ (in red), the minimising vorticity $\Omega_\star$ is lower and may even be of opposite sign. This can be seen on the right-hand panel, which shows the action-minimising vorticity $\Omega_\star$ as a function of the final time $\tf$.}}
  \label{fig:solidrotation}
\end{figure}

It is easily understood that for an infinitesimally short time lag $\tf$, there is a one-to-one correspondence between the initial velocity field and a final map that is sufficiently close to identity: $\bX_{\rm f} \approx \id +\tf\,\bv_0$. This mapping actually extends to longer time lags and a given solution to the initial-value problem can then be retrieved by solving (\ref{eq:AVP}) with the appropriate boundary conditions. This is however true only up to a finite value $t_\star$ of the time lag.
\add{In fact, the critical time $t_\star$ can be heuristically understood as a turnover time.
The reason why one-to-one correspondence breaks down for $\tf>t_\star$ becomes apparent when one considers  simple examples of vortical flows.}
Let us for instance consider a 2D solid-rotating flow in a disk of radius $R$. In Eulerian coordinates the velocity field reads
\begin{equation}
  \bv_{\Omega}(\bx,t) = \boldsymbol{\Omega} \times \bx  \;\;\; \text{with $\boldsymbol{\Omega}:=\Omega \hat \bz$  a  prescribed (constant) vorticity.}
 \end{equation}
This is a steady solution to the Euler equations. The corresponding Lagrangian flow is obviously obtained as:
\begin{equation}
  \bX_{\Omega}(\ba,t) = \ba \,\cos \Omega t + \hat \bz\times \ba \,\sin \Omega t.
  \label{eq:LagrangianSolidRotation}
\end{equation}
We can now seek to apply the boundary-value formulation (\ref{eq:AVP}) to reconstruct the rotating dynamics \add{between time $0$ and a final prescribed time ${\tf}$.} 
The final map is naturally prescribed as $\bX(\cdot,\tf) = \bX_{\Omega}(\cdot,\tf)$.  Observe that this end-point condition determines the rotation modulo $2 \pi$ only: 
\add{In other words, writing $\theta_{\rm f} \in ]-\pi,\pi]$ the principal value of the final angle $ \Omega \,\tf$,  all the solid rotations with pulsation $\Omega'_k:=  \,(\theta_{\rm f} + 2 k \pi ) /\tf$ with $k\in \mathbb Z$ share the same final map.
For such rotations, one explicitly computes Arnold's action as
\begin{equation*} 
 \mA_{0,\tf} = \dfrac{\pi R^4  }6 \Omega_k'^2 \tf = \dfrac{\pi R^4  }{6\tf} \,(\theta_{\rm f} + 2 k \pi )^2.
\end{equation*}
When $\theta_f <\pi$, the action is obviously minimal for $k=0$.  Arnold's principle therefore selects the rotation with pulsation $ \Omega_\star :=  \Omega'_{k=0} =\theta_f/\tf$, and this in turn  implies that Arnold's minimiser only coincides with the desired rotation  at pulse $\Omega$  when  $\tf<t_\star= \pi/\Omega$.  For $\tf>t_\star$, the minimiser is a solid-rotation with a smaller vorticity $|\Omega_\star|<\Omega$ (see Fig.~\ref{fig:solidrotation}). Because it solves the Euler equations, the desired solid rotation prescribed by the initial value problem still makes the action stationary but is not peculiar among the other critical points.}
\add{It is also apparent in Fig.~\ref{fig:solidrotation} and worth pointing out,  that except for the discrete set of final times   $\tf_k = k\pi/\Omega, k\in \mathbf Z$, Arnold's principle still selects a \emph{unique} rotation. While it is not the  desired rotation, Arnold's principle still can be solved univocally.
Such is not the case when $\theta_f =\pi$, which occurs  
 at times $\tf_k = k\pi/\Omega, k\in \mathbf Z$, when the final map describes a system that has rotated by exactly half of a full turn. 
 The minimal action is then achieved by both the clockwise ($k=0$) and anti-clockwise ($k=-1$) rotating flows: two minimisers exist and Arnold's principle becomes unable to select a unique minimiser.
}

%
Inducing from the solid-rotation example, it becomes intuitively clear that the presence of local rotation restricts the use of the boundary value formulation in terms of Lagrangian maps. In that sense, the critical time  $\tf$ can be tied to a notion of turnover time.\add{ When solved on time intervals larger than  this critical time, Arnold's principle (\ref{eq:AVP}) may fail to select the desired solution. In the solid-rotation case, the degeneracy relates to the final map  being compatible with  two different  pulses at the same time, that generate  equally energetic solutions. It will soon be apparent that this turnover time coincides with a critical time over which   Brenier's formulation generate genuinely probabilistic flows.}

\subsection{Brenier's least-action principle}
As previously evoked, another restriction of Arnold's description comes from its deterministic nature, even when $\tf$ is small enough. In particular, the description of a fluid in terms of a flow map cannot appropriately reproduce the ``rough'' solutions to the Euler equations that could be of relevance for inertial turbulence. \\

In Brenier's generalisation of (\ref{eq:AVP}), the fluid is no longer defined in terms of a Lagrangian flow map $\bX(\ba,t)$, but rather in terms of a \emph{generalised flow}: namely, a \emph{probability measure} $\mu[\mathfrak{D}\bZ]$ defined over Lagrangian trajectories $t\mapsto \bZ (t)$.  We only consider differentiable trajectories, so that one can define the time-integrated energy \begin{equation*}
  \mS[\bZ]= \dfrac{1}{2}\int_0^{\tf}\text{dt}\, \|\partial_t \bZ(t)\|^2.
\end{equation*} 
The Lagrangian random variable $\mS[\bZ]$ is now the fundamental ``energy functional'' quantity that generalises the action. Still, generalised flows also require generalised constraints. Essentially, we prescribe to a fixed distribution the two-time probability $\mu|_{0,\tf}$ of $t=0$ and $t=\tf$, which is obtained as a marginal of $\mu$. As for the incompressibility constraint, it translates into imposing that the one-time distributions $\mu|_t$, given again by $\mu$, stay uniform as a function of time. Brenier's least action principle can now be formally stated:
\begin{equation}
  \tag{BVP}
\begin{split}
  \mB_{0, \tf} [\mu] :=
  \int \mu[\mathfrak{D}\bZ] \,\mS[\bZ] \longrightarrow \text{inf} \qquad \text{subject to} &\\
  \mu|_{0,\tf}= \mu_{0,{\rm f}} \;\;\text{and} \;\; \mu|_t \mbox{ uniform}.&
\end{split}
  \label{eq:BVP}
\end{equation}
The probabilistic constraints are on marginals of the distribution $\mu$, whence the relationship with multi-marginal optimal transport.  Because there is a probabilistic coupling $\mu_{0,{\rm f}}$  between the initial and final positions, the formulation (\ref{eq:BVP}) is again a boundary-value problem.\\

The generalised least-action principle (\ref{eq:BVP}) was conceptualised by \cite{brenier1989least}\,---\,see also \cite{brenier2008generalized}\,---, with some established mathematical properties, including:
\begin{enumerate} 
\item Generalised solutions to (\ref{eq:BVP}) exist for very general end-point couplings $\mu_{0,{\rm f}}$. \item The generalised variational principle encompasses classical (regular) solutions, provided that the final time $\tf$ is small enough. This requires choosing a deterministic end-point coupling, which is prescribed by the map associated to a solution to the Euler equations, \textit{i.e.} $\mu_{0,{\rm f}}(\mathrm{d}\ba,\mathrm{d}\bx) = |\mD|\,^{-1}\delta(\bX_{\rm f}(\ba)-\bx)\,\mathrm{d}\ba\,\mathrm{d}\bx $. The solution to (\ref{eq:BVP}) is then highly degenerate, concentrated on the deterministic Lagrangian flow. This requires $\tf<t_\star$ with
  \begin{equation}
    t_\star  = \dfrac{\pi}{\sup_{\bx,t}\| \text{Hessian}(p)\|^{1/2}},
    \label{eq:breniertime}
  \end{equation}
  with $p$ denoting the pressure field appearing in the underlying Euler equations.
  \item  For $\tf>t_\star$, generalised flows can become non-deterministic: The flow is no more described by the Lagrangian map, but rather by transition probabilities. A generalised ``fluid particle'' tagged by its initial position then splits mass and have a multiplicity of subsequent positions at time $t$ prescribed by the two-time marginals $\mu|_{0,t}$.
\end{enumerate}

Point (ii) above shows that (\ref{eq:BVP}) is indeed a generalisation of (\ref{eq:AVP}). Yet, the short-time restriction of Arnold's boundary-value formulation remains: Seemingly, considering $\tf>t_\star$ might not be of physical relevance. Note in passing that for a solid rotation the pressure field is \add{$p(x,y)= ({\Omega}/{2})^2(x^2+y^2)$}, yielding in that case $t_\star = \pi/|\Omega|$ as obtained in \S\ref{ssec:arnold-solid}.  In fact, the ``generalised'' aspect of (\ref{eq:BVP}) relates to the type of Euler solutions that it can construct. Those can in principle include weak solutions to the Euler equations, be them distributional or even measure-valued \citep{brenier1999minimal}. In those cases, the generalised flow is always non-deterministic, even at short $\tf$. 

In addition to those conceptual features, Brenier's formulation has also a practical outcome~: While (\ref{eq:AVP}) is a highly non-linear problem which is cumbersome to use in practice, (\ref{eq:BVP}) is a linear optimisation problem, \add{ that is, a problem in which both the objective function and the constraints depend linearly on the arguments to be minimised upon.
Various  numerical methods are known, that can efficiently handle  either low-dimensional or  high-dimensional linear searches \citep{nocedal2006numerical,benamou2017generalized,merigot2016minimal,gallouet2016lagrangian}. 
}  
The relevance of generalised flows when used in combination with turbulent inputs can therefore be tested, and this remarkable feature provides the guideline for the remainder of the paper.


\section{Statistical physics of generalised flows}
\label{sec:statphys}
The generalised variational principle (\ref{eq:BVP}) connects inviscid fluid mechanics with multimarginal optimal transport. Its numerical integration can thus benefit from the efficient algorithms that have been developed to solve such optimisation problems. In recent years, (\ref{eq:BVP}) has even been used as a testground in the  development of  such methods\,---\,see for instance \cite{brenier2008generalized,benamou2015iterative,benamou2017generalized,nenna2016numerical,gallouet2016lagrangian}.
Here we follow a different strategy. We rather rely on a Monte-Carlo algorithm, as the latter proves sufficient for our present scope. It stems from an intuitive statistical-mechanics interpretation of (\ref{eq:BVP}). 
\subsection{Coarse-graining and  permutation flows}
\begin{figure}
  \centerline{\includegraphics[width=0.7\textwidth]{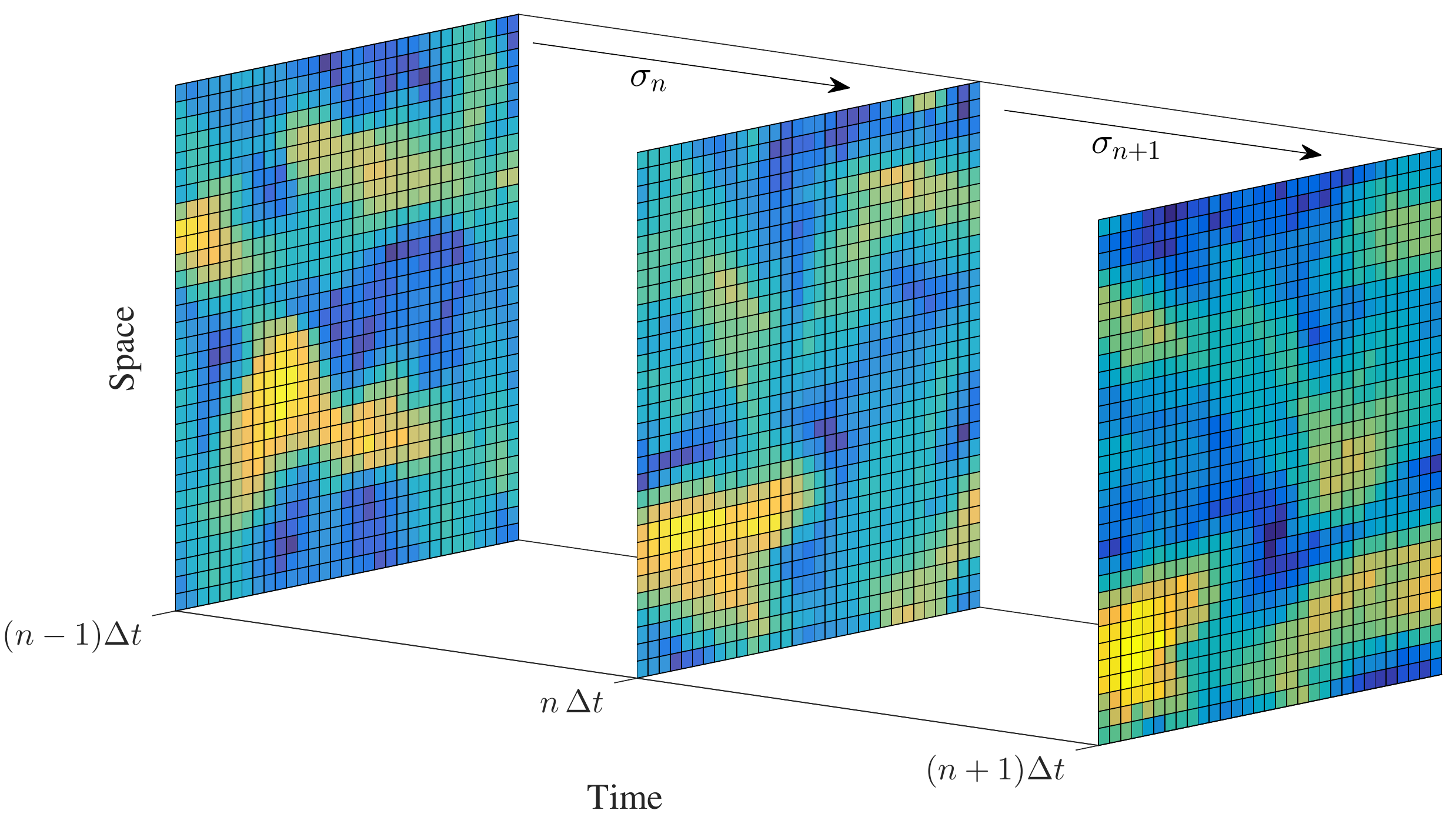}}
  \caption{Illustration of \add{a realisation} of two-dimensional permutation flow. Colors code the initial positions.}
  \label{fig:permutation}
\end{figure}

It is illuminating to reformulate (\ref{eq:BVP}) in a discrete setting, both in space and in time. 
Specifically, we discretise time into $N_t$ steps $t_0=0,t_1, \dots, t_{N_t}=\tf$, and 
partition the $d$-dimensional physical domain $\mD$ into $N_x^d$ ``boxes'' of equal volume, and labeled by the $d$-dimensional indices $\bi = (i_1,\dots,i_d) \in [1,N_x]^d$.
In this  ``coarse-grained'' description, Lagrangian trajectories become sequences of indices $\{ \bi_n \}_{n=0..N_t}$, and generalised flows become finite-dimensional tensors, represented by the non-negative $({N_x^d})^{N_t}$ entries $\mu\{ \bi_n \}:= \mu(\bi_0,\bi_1,\dots,\bi_{N_t})$ normalised to $1$.
One can now think of an \emph{incompressible} generalised flow in two equivalent ways~: 
\begin{enumerate}
\item In the ``Linear-programming viewpoint'', it is a measure with uniform single-time marginals, that is
  \begin{equation}
    \mu_k(\bi_k) := \!\!\!\!\!\!\!\sum_{\bi_0\dots \bi_{k-1},\bi_{k+1}\dots \bi_{N_t}} \!\!\!\!\!\!\!\mu \{ \bi_n \} = 1/N_x^d,  \quad\mbox{for all } k=0..N_t.
  \label{eq:singlemarg}
  \end{equation}

\add{In this discrete setting, the optimisation problem (\ref{eq:BVP})  consists in minimising a suitably discretised average linear energy functional $\mu \mapsto \sum_{\bi_n} \mu \{ \bi_n \}  \mathcal E_d\{ \bi_n \}$,  over the  non-negative tensors $\mu$ satisfying the single-time marginal constraints  (\ref{eq:singlemarg}), and  two-time bounday coupling prescibed in the form of a doubly stochastic matrix, assigning a transition probability from box ${\bf i}$ to box ${\bf j}$ between initial and final time. In this formulation, it is apparent  that the fluid parcel in box ${\bf i}$ at initial time is allowed to split mass. This is natural provided that  one interpretes the  generalised flow as a spatial coarse-graining of the Lagrangian trajectories.  Taking the joint limit ${N_t\to \infty,N_x\to\infty}$ one formally retrieves the continuous formulation (\ref{eq:BVP}).
We refer the interested reader to the pedagogic review chapters of  \citep{nenna2016numerical} for precise mathematical statements describing the convergence of discrete solutions towards  time-continuous minimising flows.
}

\item In the ``Statistical mechanics viewpoint'',  it is a measure prescribed  by a certain ensemble average over \emph{``permutation flows''} $\bsigma$ \citep{brenier2008generalized}, that is
  \begin{equation}
    \mu\{ \bi_n \} \equiv \dfrac{1}{N_x^d}\left\langle \prod_{k=1}^{N_t} \delta_{\bi_k\bsigma_k(\bi_0)} \right\rangle_\bsigma
  \end{equation}
\end{enumerate}
In the previous formula, each ${\bsigma}_n$ is a permutation over the $N_x^d$ boxes between step  $0$ and step $n$. Taking $\bsigma_{0}=\id$, the sequence $\{ \bsigma_n(\bi) \}_{n=0..N_t}$ then represents the coarse-graining of a Lagrangian trajectory starting from box $\bi$, and the sequence $\bsigma = \{ \bsigma_n \}_{n=0..N_t}$ is termed the ``permutation flow'' (see illustration in Fig.~\ref{fig:permutation}).  With this second interpretation at hand,  we can now reformulate (\ref{eq:BVP}) in terms of an optimal statistical ensemble of permutation flows.
\subsection{Generalised solutions as statistical ensembles of permutation flows.}
\label{subsec:permutations}
Each realisation of a permutation flow yields the discrete action~:
\begin{equation}
  \mA_{0, N_t}[\bsigma]= \sum_{n=1}^{N_t}
  \mathcal{E}_n\quad \mbox{with}\quad  \mathcal{E}_n = 
  \sum_{\boldsymbol{i}} \|  \boldsymbol{\sigma}_n (\boldsymbol{i}) -
  \boldsymbol{\sigma}_{n-1} (\boldsymbol{i}) \|^2,
  \label{eq:DiscreteAction}
\end{equation}
where $\|\bi-\bj \|$ denotes here the discrete distance in $\mathbb{Z}^d$ between the centres of boxes $\bi$ and $\bj$.
We now formulate the discrete counterpart to (\ref{eq:BVP}) as an optimisation over discrete measures $p[\bsigma]$ defining statistical ensembles of permutation flows:
\begin{equation}
  \tag{BDVP}
    \mB_{0,N_t} \left[p\right] :=
  \sum_{\bsigma} \mA_{0, N_t}\left[\bsigma\right] \,p[\bsigma] \longrightarrow \text{inf} \quad \text{subject to}\ \
 \sum_{\bsigma}\bsigma_{N_t}\,p[\bsigma] = \mu_{0 \to\tf},
  \label{eq:BDVP}
\end{equation}
where  $\mu_{0 \to\tf}$ is the transition probability from initial to final time. Using the standard matrix representation of permutations, $\mu_{0 \to\tf}$ is an $N_x^d \times N_x^d$ doubly stochastic matrix. The discrete formulation (\ref{eq:BDVP}) reveals that Brenier's formulation resembles an averaged Arnold's principle, whereby averages are taken with respect to measure-preserving applications, a set which includes but is not restricted to smooth volume-preserving maps.

The optimal statistical measure that solves (\ref{eq:BDVP}) is discrete, but its dimensionality is huge\footnote{explicitly : ${(N_x^d)!} ^{N_t}$.}! Monte-Carlo algorithms are thus particularly suited to obtain numerical estimates of averages with respect to the optimal flow.  Specifically, the formulation (\ref{eq:BDVP}) can be formally understood as the zero-temperature/infinite-$\beta$ limit of the Gibbs canonical measure
 \begin{equation}
  p_\beta[\boldsymbol{\sigma}] = \frac{1}{Z(\beta)}\mathrm{e}^{-\beta\, \mathcal{A}_{0,N_t}[\boldsymbol{\sigma}]} \;\;\text{with associated averages } \;\;\langle \cdot \rangle_\beta.
  \label{eq:gibbs}
\end{equation}
For finite temperatures, Monte-Carlo averages are then easily sampled using a local Metropolis algorithm \citep[see, e.g.,][]{binder1986introduction} on a $(d+1)$-dimensional lattice, with the extra dimension representing the time axis. 
\add{Algorithm \ref{algorithm:mc} provides the  pseudo-code of  a single Monte-Carlo iteration in our two-dimensionnal setting. }
We note that the finite-temperature regularisation is probably a statistical-mechanics counterpart to the concept of entropic regularisation used by \cite{nenna2016numerical} to implement efficient iterative projection algorithms rooted in the ``Linear programming viewpoint''.
\add{As a side remark, one can also point out  that a realisation of a permutation flow, as represented in Fig.~\ref{fig:permutation} and defined by the energy functional   (\ref{eq:DiscreteAction}) , may as well be  thought of as a configuration of a generalized Potts model  with $N_x^d$ colors and anisotropic nearest-neighbour interactions (see \emph{e.g.} \cite{baxter2016exactly}).}

\begin{algorithm}[h]
\linespread{1.4}\selectfont
\alglanguage{pseudocode}
\caption{Pseudo code of the  Monte-Carlo iteration used to estimate the statistical ensemble (\ref{eq:gibbs}). Permutations flows are here represented in terms of a 2d-integer array $\bsigma = (\sigma_{\bi t})_{\bi \in N_x^2,t\in N_t}$.}
\label{algorithm:mc}
\begin{algorithmic}[1]
\Procedure{Monte-Carlo Iteration}{}	
     \State Pick random  $(\bi,\bj)  \in [|1,N_x^2|]^2$ and $t  \in [|2,N_t-1|]$.
      \Comment{Candidate Switch}
     \State {$ \Delta \mathcal E \gets \sum_\pm \left( \|\sigma_{\bj t}-\sigma_{\bi  t\pm1} \|^2+\|\sigma_{\bi t}-\sigma_{\bj  t\pm1} \|^2\right)$ \newline
\hspace*{5em}$-\sum_\pm \left(\|\sigma_{\bi t}-\sigma_{\bi  t\pm1} \|^2-\|\sigma_{\bj t}-\sigma_{\bj  t\pm1} \|^2\right)$}
      \Comment{Energy Estimate}
     \State Pick random $u \in [0,1]$
	\If {$u < e^{-\beta \Delta \mathcal E}$}      \Comment{Metropolis rejection rule}
		\State Switch $\sigma_{\bj t}$ and $\sigma_{\bi t}$
       \EndIf

\EndProcedure
\end{algorithmic}
\end{algorithm}

\subsection{Large-time limit of optimal permutation flows.}
\label{sec:Fourcell}
\begin{figure}
  \centering
	\begin{minipage}{0.27\textwidth}
		\includegraphics[width=\textwidth,page=2]{FourCell_Redux}\\
		\includegraphics[width=\textwidth,trim=0.5cm 0cm 0.5cm 0.5cm,clip]{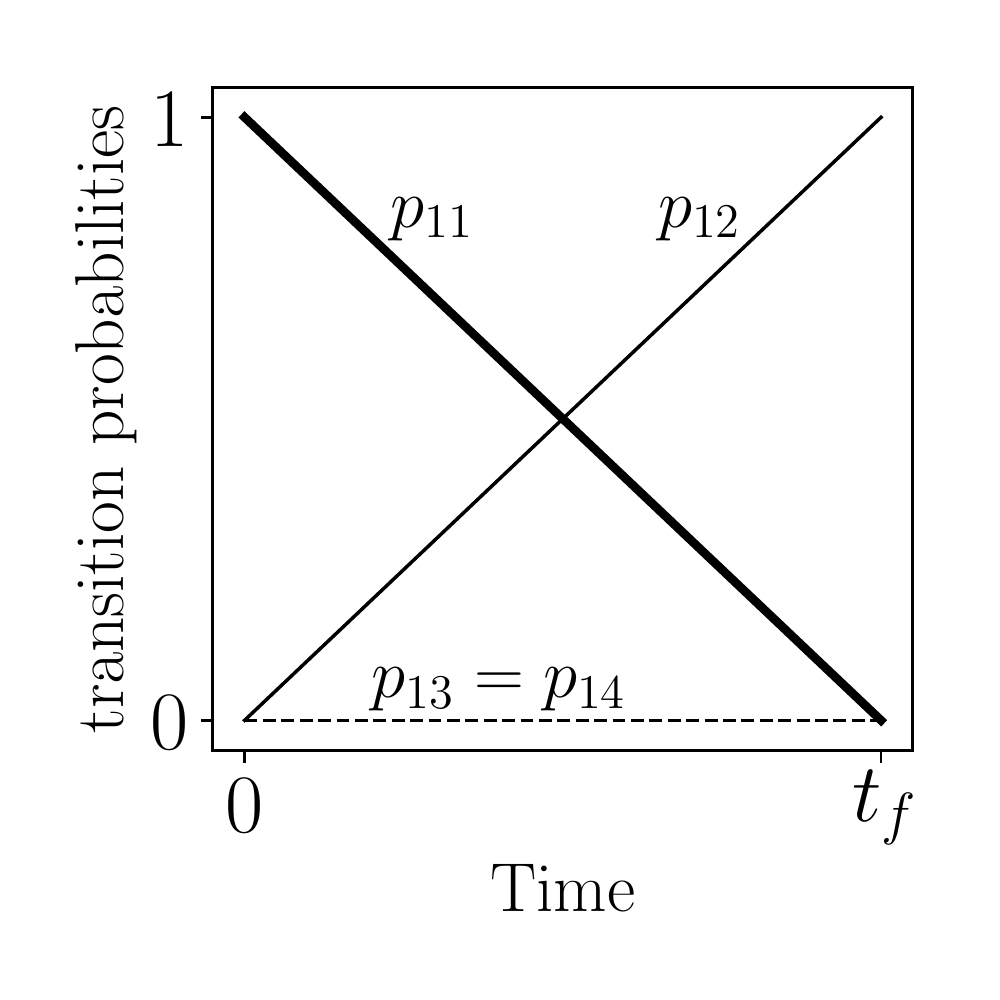}
	\end{minipage}
	\begin{minipage}{0.19\textwidth}
		\centering
		\includegraphics[width=0.6\textwidth,page=1]{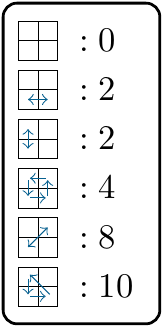}
		\vspace{3cm}
	\end{minipage}
	\begin{minipage}{0.27\textwidth}
		\includegraphics[width=\textwidth,page=3]{FourCell_Redux}\\
		\includegraphics[width=\textwidth,trim=0.5cm 0cm 0.5cm 0.5cm,clip]{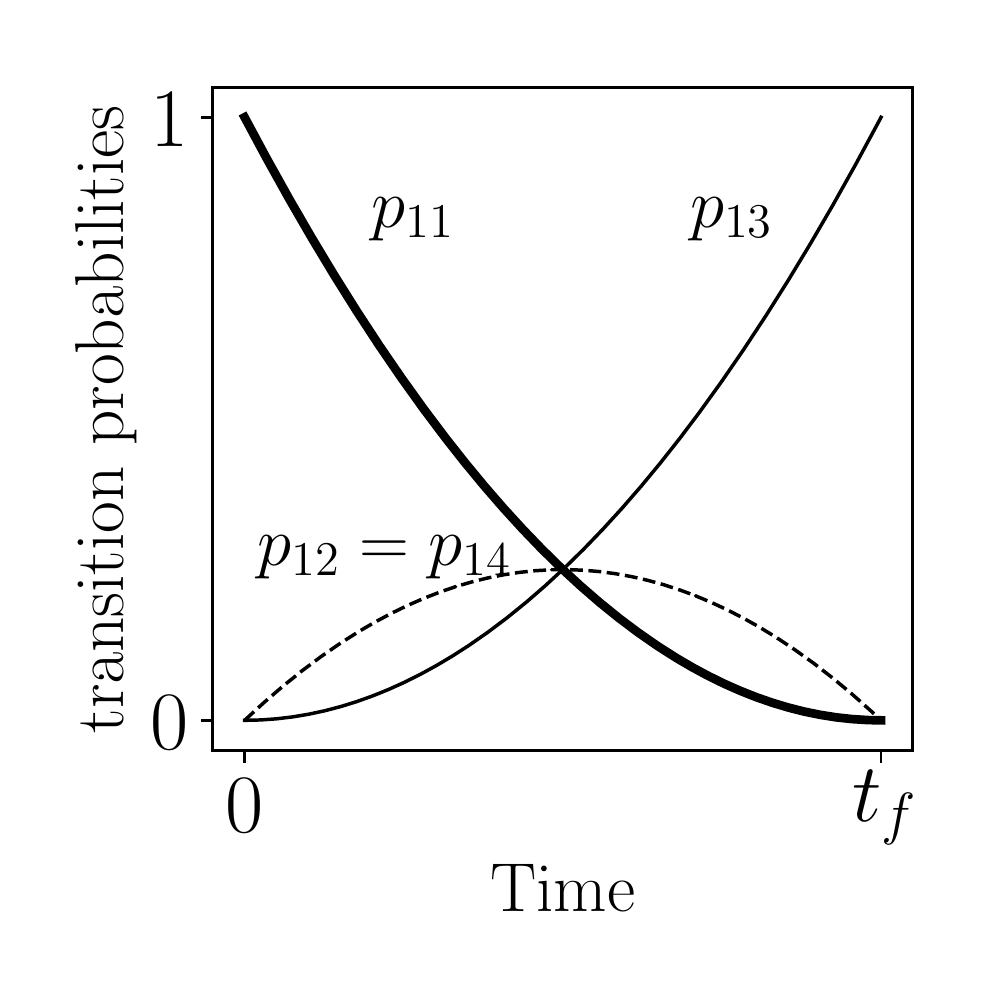}
	\end{minipage}
	\caption{\add{The solid rotation example, this time interpreted as an ensemble of permutation flows on a coarse $(2\times2)$-box: nothing but a sliding puzzle. The middle inset provides the energy cost of the representative elementary permutations: Permutations flows are discrete sequences of $N_t$ such elementary permutations,  and the minimising discrete generalised flow is defined as the  uniform probability measure over the set of permutation flows with minimal energy cost.
		The top panel represents the minimal permutation flows, when the final permutation is either a quarter of a full turn (left side) or half of a full turn (right side).
		The physical trajectory is uniquely defined in the first case as  an anti-clockwise rotation. In the second case, the generalised flow is an average over both clockwise and anti-clockwise half-turn rotations.
		The bottom panel represents the time evolution of   Lagrangian transition probabilities  for the corresponding generalised flow, obtained in the limit $N_t \to \infty$ (see text for details).}}
	\label{eq:Fourcell}
\end{figure}

\add{
One apparent virtue of (\ref{eq:BDVP}) lies in its  ability to produce unique generalised minimisers, when Arnold's approach can fail to select one.
From the perspective of discrete permutation flows,  one can  however hint that the uniqueness of such generalised minimisers (\ref{eq:BDVP}) ressembles a mathematical artefact rather than a new physical solution.
Let us for instance  revisit the solid rotation example of \S\ref{ssec:arnold-solid}, in terms of a very coarse generalised flow involving a $(2\times 2)$-spatial discretisation and $N_t$ time steps of length $\Delta t :=t_f/N_t$.
In the solid-rotation example, and as was apparent in Fig.~\ref{fig:solidrotation}, we recall that Arnold's principle only failed to select a solution when the final map represents half of a complete turn, in which case both clockwise and anti-clockwise rotations were equally energetic.
Figure~\ref{eq:Fourcell} illustrates how the permutation flow viewpoint accommodates this degeneracy.
When the final state is say an anti-clockwise quarter of a full turn, the  permutations flows with minimal costs are the sequences of  permutations $\sigma_1 \cdots \sigma_{N_t}$, where except for  one non-trivial $\sigma_t := \circlearrowright$  representing an anti-clockwise 4-cycle, every other one is the identity. 
The generalised minimiser is then obtained as the uniform average over those permutation flows, and the Lagrangian trajectories $\bX(t,\bi)$ starting at prescribed $\bi$ are then uniquely defined in physical space. Using elementary combinatorics, one can compute the transition probabilities $p_{\bi\bj}(t):=\mu(\bX(t,\bi)=\bj)$ and obtain in the limit  $N_t \to \infty$,
$p_{13}=p_{14} = 0$, together with $p_{11}=1-p_{14}=1-t/t_f$. Rightfully, the generalised solution predicts that no particle  starting in the top-left box  $\bi_0=1$ transits diagonally or to its left before going down, and this generalised dynamics is represented in the left panel of Fig.~\ref{eq:Fourcell}.\\
We can now consider the case, when the final state is half of a full turn, and this situation is shown in the right panel  of Fig.~\ref{eq:Fourcell}.   The  permutations flows with minimal costs are now the sequences of permutations, containing  exactly two non-trivial 4-cycle   at arbitrary times $(t_1, t_2) \in [|1,N_t|]^2$, that is  $\sigma_{t_1} =\sigma_{t_2}:= \circlearrowright$ or $\sigma_{t_1} =\sigma_{t_2}:= \circlearrowleft$. The Lagrangian trajectories are then non-uniquely defined but the generalised solution is so. In the limit  $N_t \to \infty$, one for instance obtains the following transition probabilities:
$p_{12}=p_{14} = (t/\tf)(1-t/\tf)$, and $p_{11}=(1-t/\tf)^2$, $p_{13}=(t/t_f)^2$.
Particles contained in the top-left box have now  equal probability to either clockwise or anti-clockwise transit towards the bottom-right corner.\\
It is  apparent that the mass-splitting is somewhat artificial: It solves the degeneracy of the deterministic problem by super-imposing the possible trajectories and allowing the particles to feel both clockwise and anti-clockwise rotation at the same time.
This toy argument suggests that there could be no physical meaning for the generalised variational formulation when Arnold's principle breaks down.  With that hint in mind, we can now study the physical properties of more intricate generalised flows, for which the vorticity field is no longer a mere constant of space and time, and the final map no longer necessarily degenerate.
}

\section{Generalised flows and  Beltrami dynamics}
\label{sec:Beltrami}

Beltrami flows are a class of strong Euler solutions of next-order complexity compared to solid rotations~: While stationary, their vorticity is not constant in space. 
It is therefore instructive to study the statistical features of the generalised flows induced when using Beltrami Lagrangian map as boundary conditions.
Using our Monte-Carlo algorithm,  we find that the generalised flows indeed reproduce the intermediate Lagrangian dynamics of Beltrami flows for small times. For large final times, the Lagrangian flow breaks down. Even if this could have a promising ``turbulent flavour'' and provide a possible framework to model intrinsic stochasticity, a deeper analysis of their statistical features reveals that such generalised flow are actually unphysical.
\subsection{Generalised flows from Beltrami maps}
We consider the one-cell Beltrami flow defined by the Eulerian velocity field on the domain $\mD= [0,1]^2$:
\begin{equation}
  \bv(\bx,t) =\left( \cos \pi y \,\sin \pi x,  -\cos \pi x \,\sin \pi y \right)^{\mathsf{T}},
\end{equation}
from which we obtain explicitly $t_\star =1$ using Brenier's formula (\ref{eq:breniertime}).
\begin{figure}
  \begin{center}
    \includegraphics[width=\textwidth]{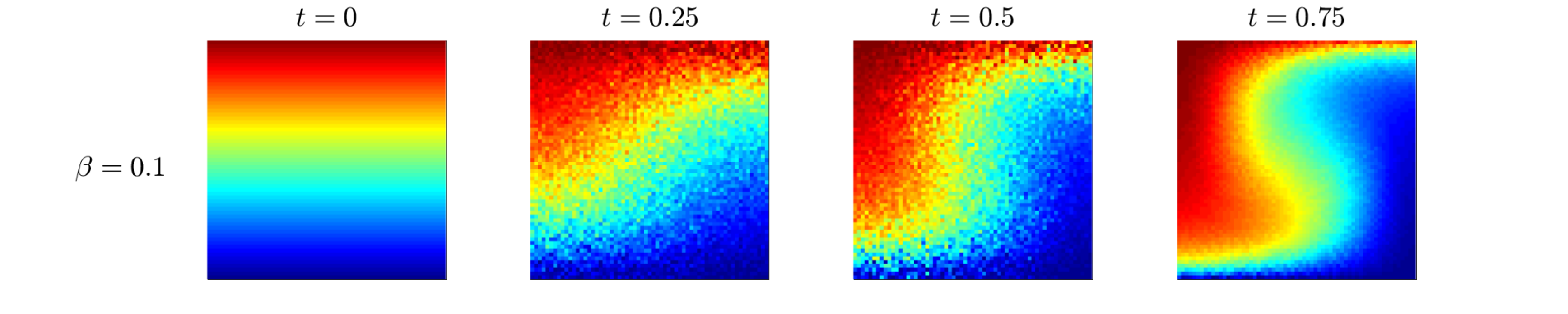}\\
    \includegraphics[width=\textwidth]{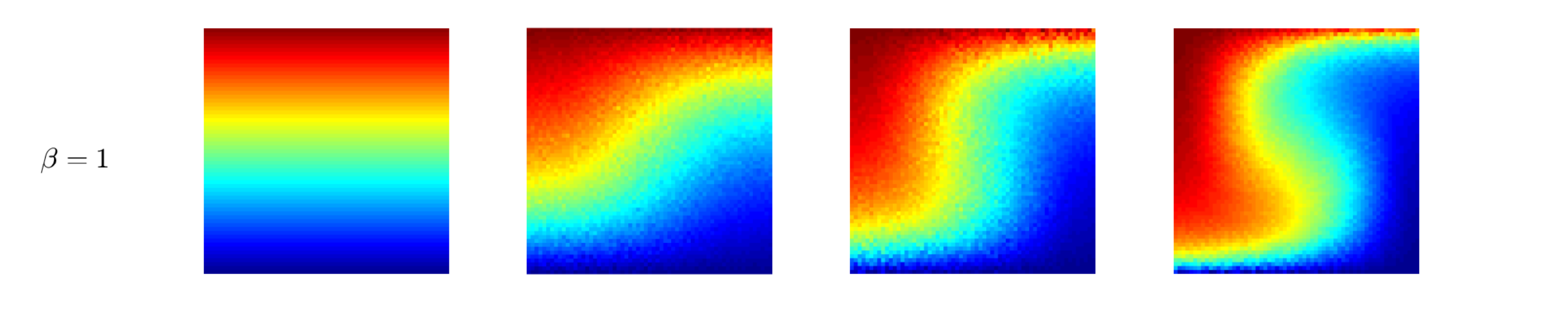}\\
    \includegraphics[width=\textwidth]{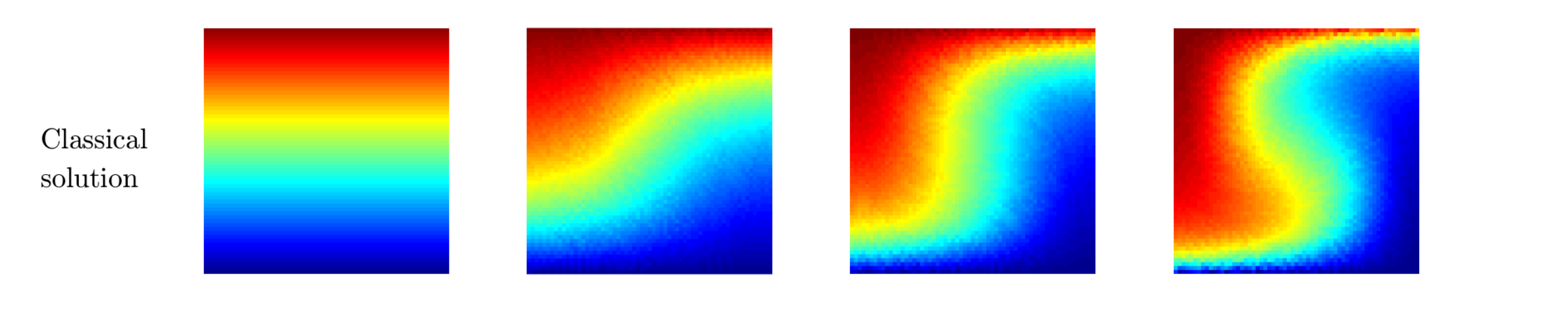}
  \end{center}
  \caption{Reconstruction of the Beltrami dynamics as a function of time and $\beta$ for $\tf = 0.75 < t_\star$. The colour code stands for the initial $y$ position of fluid particles.}
  \label{fig:BeltramiEvo}
\end{figure}
Tracer particles are used to generate Beltrami permutation maps on a $64^2$ spatial grid at different prescribed final times. The generalised flows are then constructed with fixed time-step $ \Delta t= t_\star/8$. For instance, when $\tf = t_\star$,  the discretisation has parameters  $N_x=64$ and $N_t = 8$ in the notations of \S\ref{sec:statphys}. To ensure that a statistical steady state is attained, averages are computed after routinely discarding the first $10^6$ Monte-Carlo times\footnote{One Monte-Carlo time is standardly defined as $N_x^2\times N_t$ iterations of the algorithm.}, and then using the data generated over the next $2.5\times10^{5}$ ones. Our numerics clearly show  that $t_\star=1$ marks a transition between a deterministic and a stochastic regime for the generalised flow.

\subsection{$\tf<t_\star$ : Convergence toward Beltrami dynamics.}
\begin{figure}
  \centerline{\includegraphics[height=0.45\textwidth]{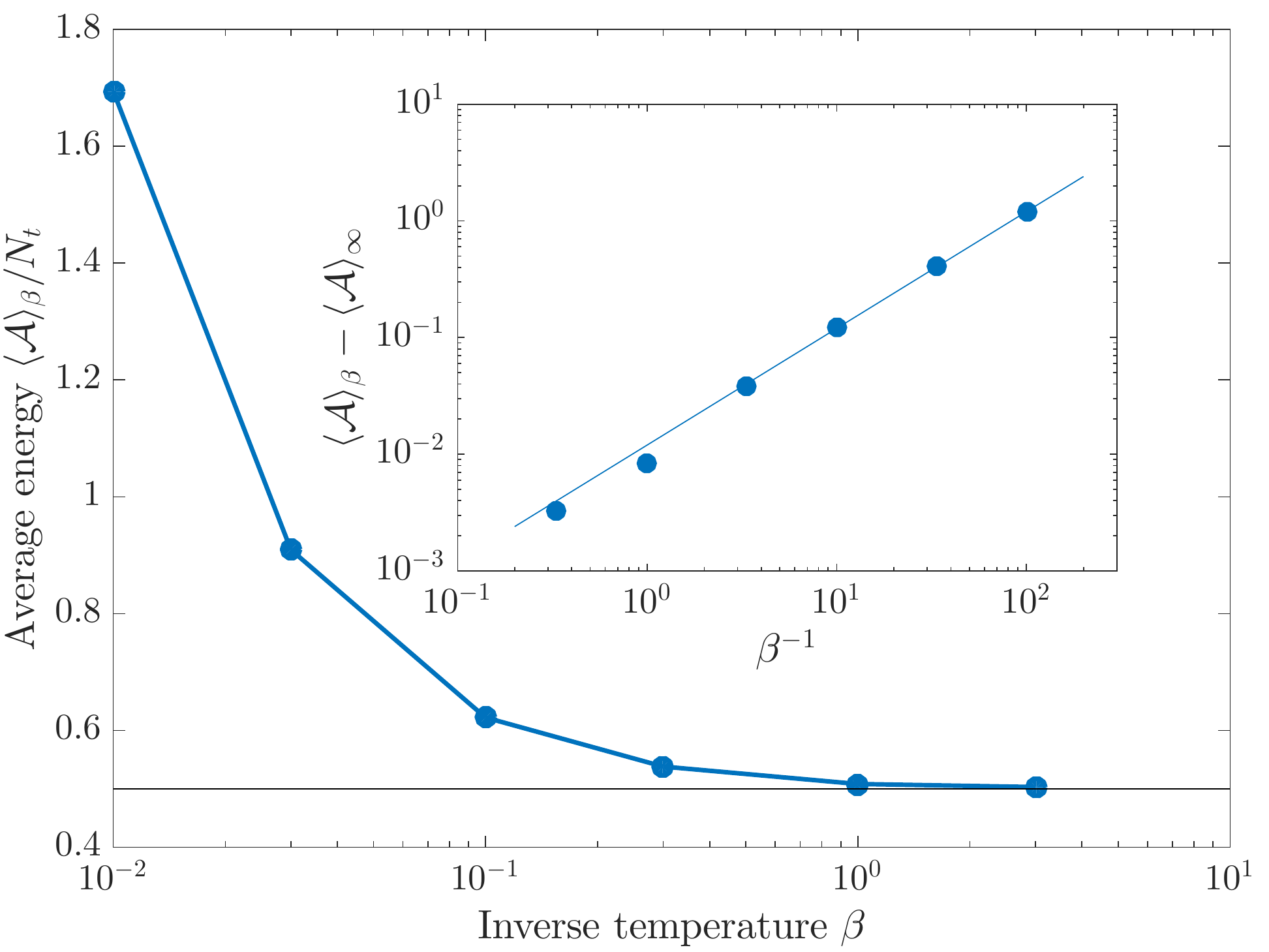}
    \qquad \includegraphics[height=0.45\textwidth]{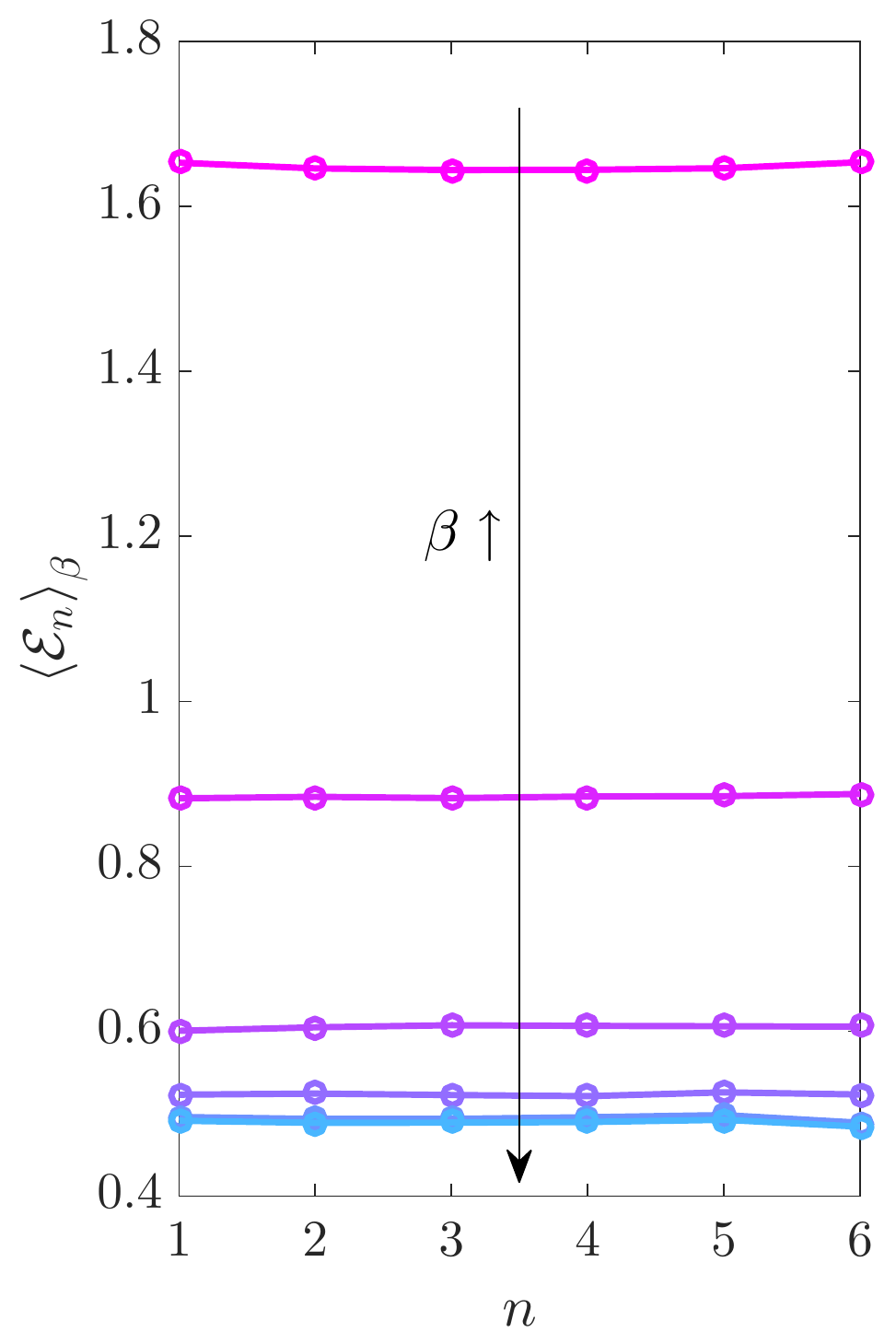}}
  \caption{Convergence of the action for the case $\tf < t_\star$. The left-hand panel shows convergence at zero temperature of the time-averaged kinetic energy\,---\,given by the action in (\ref{eq:BDVP})\,---\,towards the optimal Beltrami action. The right-hand panel shows the time dependence of the generalised flow energy for inverse temperatures $\beta=0.01$, $0.03$, $0.1$, $0.3$, $1$ and $3$ (from top to bottom). }
  \label{fig:ConvergenceAction}
\end{figure}
\begin{figure}
  \centerline{\includegraphics[width=0.49\textwidth]{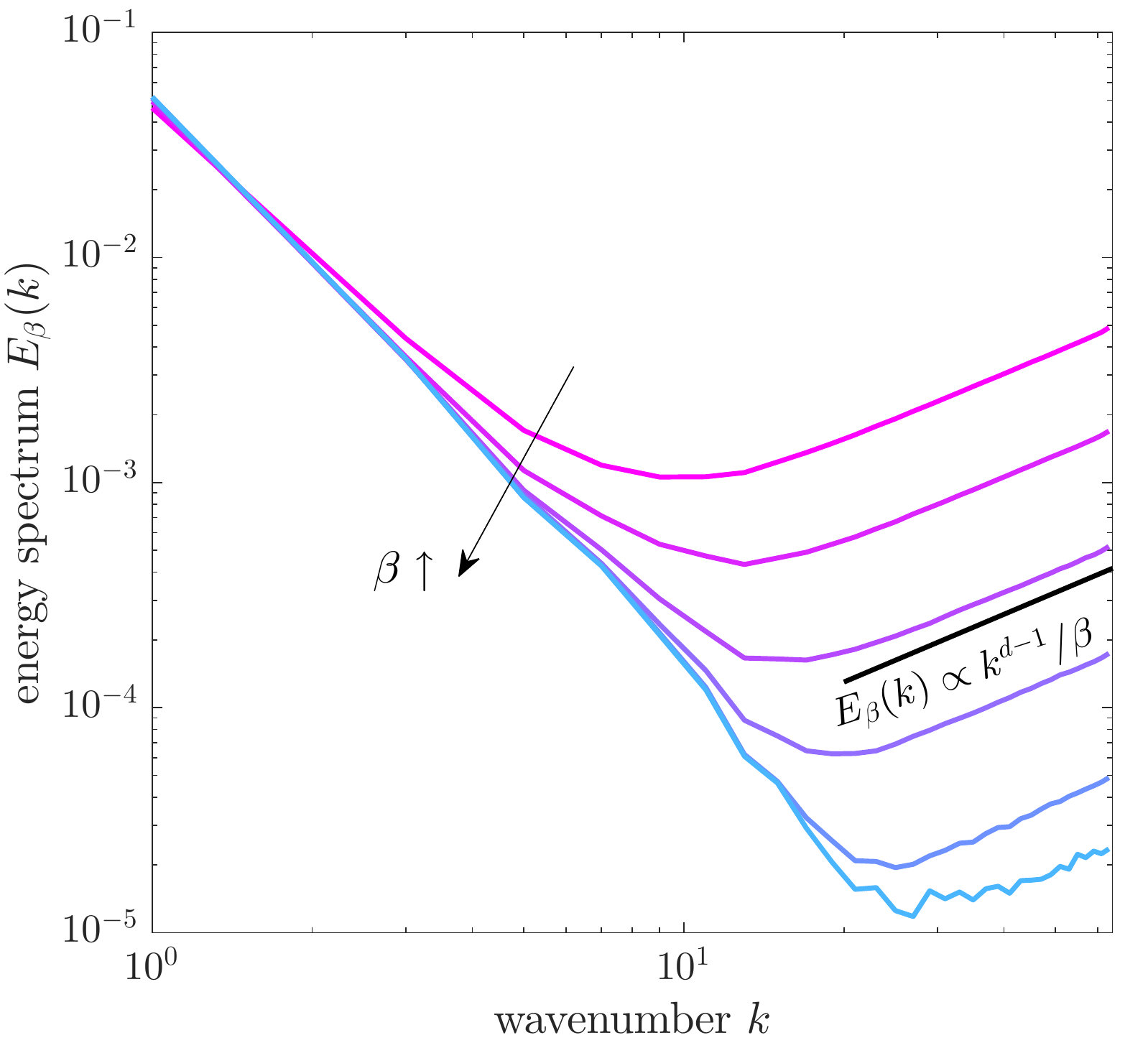}
     \includegraphics[width=0.49\textwidth]{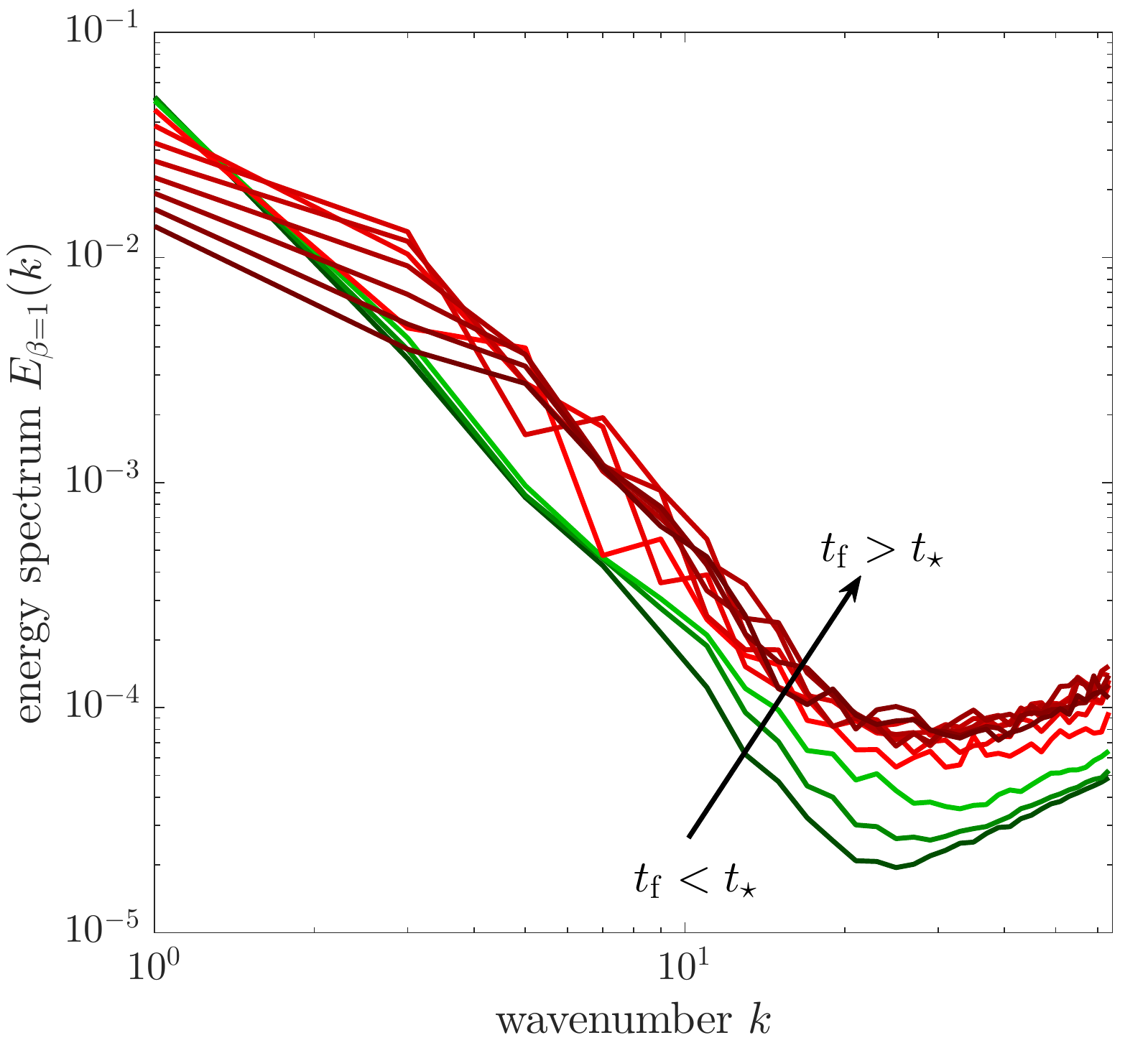}}
   \caption{Kinetic energy spectra for, on the left-hand panel, $\tf < t_\star$ and various $\beta$ (same values as Fig.~\ref{fig:ConvergenceAction} left) and, on the right-hand panel, for $\beta=1$ fixed and various $\tf$ ranging from $0.75$ up to $2$ every $0.125$. The curves corresponding to $\tf<t_\star$ are shown in green, the others in red. }
  \label{fig:beltramispectra}
\end{figure}

When $\tf<t_\star$, it is found that the Gibbs measures $\langle\cdot \rangle_\beta$ concentrate around the deterministic dynamics in the limit of zero temperature, $\beta \to \infty$. This can be seen in Fig.~\ref{fig:BeltramiEvo}: For small-enough temperatures, the average Lagrangian maps become essentially undistinguishable from those associated to the deterministic Beltrami flows, up to some small remaining thermal noise.  This convergence towards the optimal ensemble is shown more clearly in Fig.~\ref{fig:ConvergenceAction}: As $\beta$ increases, the average action converges towards the Beltrami action, which, as we know, solves (\ref{eq:BDVP}) when $\tf<t_\star$. As shown in inset, the convergence is algebraic $\propto \beta^{-1}$. Moreover, it is apparent from the right-hand panel of Fig.~\ref{fig:ConvergenceAction} that energy is conserved in time. This is a signature of the high regularity of such a classical solution to the Euler equations. This observed regularity is a consequence of the variational principle and was definitely not prescribed. This shows the effectiveness of Brenier's variational principle to accomodate strong, classical, deterministic solutions, in spite of its probabilistic nature. Finally, let us note that the ``finite temperature effects'' are apparent on the spectra of the displacement and translate into small-scale thermalisation, namely equipartition of energy (see the left-hand panel of Fig.~\ref{fig:beltramispectra}). The range of thermalised scales naturally decreases with increasing $\beta$. Those features show the roundabout behaviour of our numerical procedure. They illustrate the established fact that for small enough $\tf<t_\star$ the generalised formulation (\ref{eq:BVP}) do indeed reproduce the Arnold's minimisers of (\ref{eq:AVP}) when those exist, as proven in \cite{brenier1989least}.
\subsection{$\tf>t_\star$ : Non-physical behaviour of the generalised flow.}
\begin{figure}
  \centerline{\includegraphics[width=1.2\textwidth]{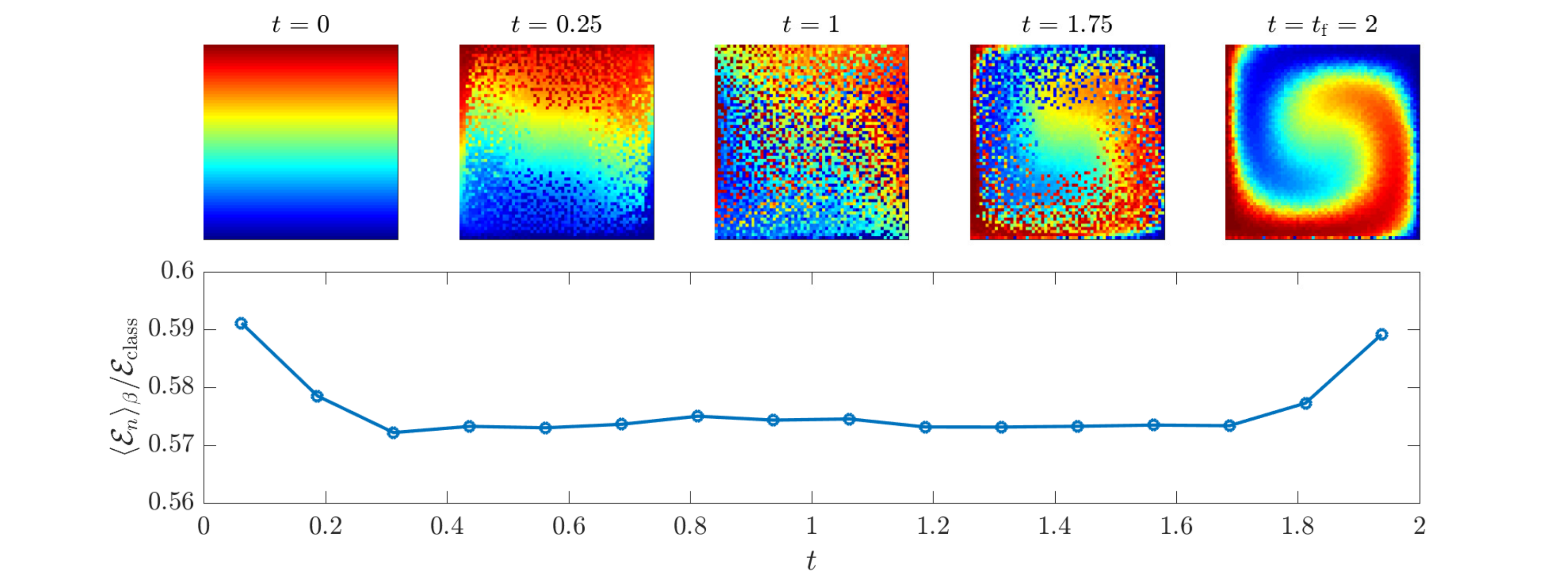}}
  \caption{Upper panel: typical permutation flow for the  generalised solution  with $\tf=2t_\star$. We again show here, as a coloured background, the initial $y$ label as a function of time. Lower panel: time-evolution of the kinetic energy. It is no longer conserved, and exhibits strong variations at the time boundaries.}
  \label{fig:BeltramiTransition}
\end{figure}

When $\tf>t_\star$, the behaviour of the generalised flow becomes non-deterministic, and the time-conservation of energy breaks down\,---\,see Fig.~\ref{fig:BeltramiTransition}. This implies in particular that generalised flows cannot solve the Euler equations in a strong sense anymore. Yet, it is also apparent that the flow is not  \emph{dissipative}: The variations in the energy profile shown in the lower panel are symmetric and localised at the boundaries. Those behaviours are non-physical and  likely  consequences of both the time-symmetry and the boundary-value formulation of the principle (\ref{eq:BVP}) itself. 
In the example chosen for Fig.~\ref{fig:BeltramiTransition},  the breakdown of the Lagrangian flow is particularly spectacular for $t=\tf/2=1$. The nearly-uniform colour pattern shows that initially distant pockets of fluid become arbitrarily close to each other at intermediate times before separating again. This is the signature of trajectory crossings. This picture may misleadingly suggest that generalised solutions essentially produce a thermalised dynamics, with fluid motion occuring at random. This is however not the case, as revealed by looking at the Eulerian spectra of the generalised displacement that are displayed in the right-hand panel of Fig.~\ref{fig:beltramispectra}. 
While the small scales are indeed thermalised,  the thermalisation is rather likely a spurious finite-temperature effect due to our numerical procedure, as its spread in $k$-space is not much wider than in the $\tf<t_\star$ case. The generalised flows obtained for $\tf>t_\star$ however display  specific large-scale dynamics, which carry far less energy than the deterministic counterparts, whence their selections as optimal flows.
We also note that, apart in the spurious thermalised range, the specific scalings  displayed by the Beltrami generalised flows have no known counterparts in the phenomenology of two-dimensional turbulence. 

\begin{figure}
  \begin{center}
  \begin{minipage}{0.55\textwidth}
    \includegraphics[width=\textwidth]{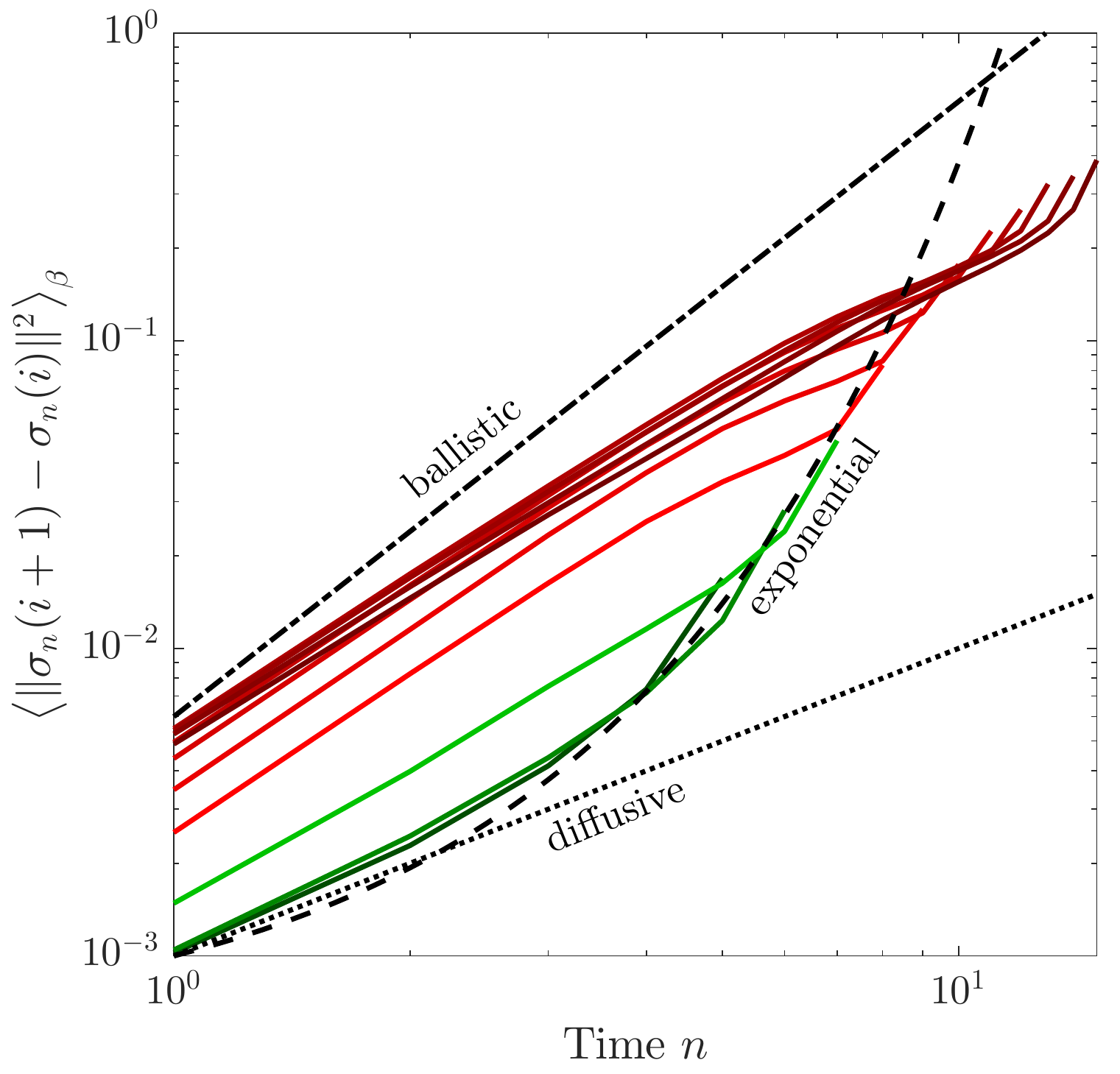} 
  \end{minipage}
  \qquad
  \begin{minipage}{0.22\textwidth}
    \includegraphics[width=\textwidth]{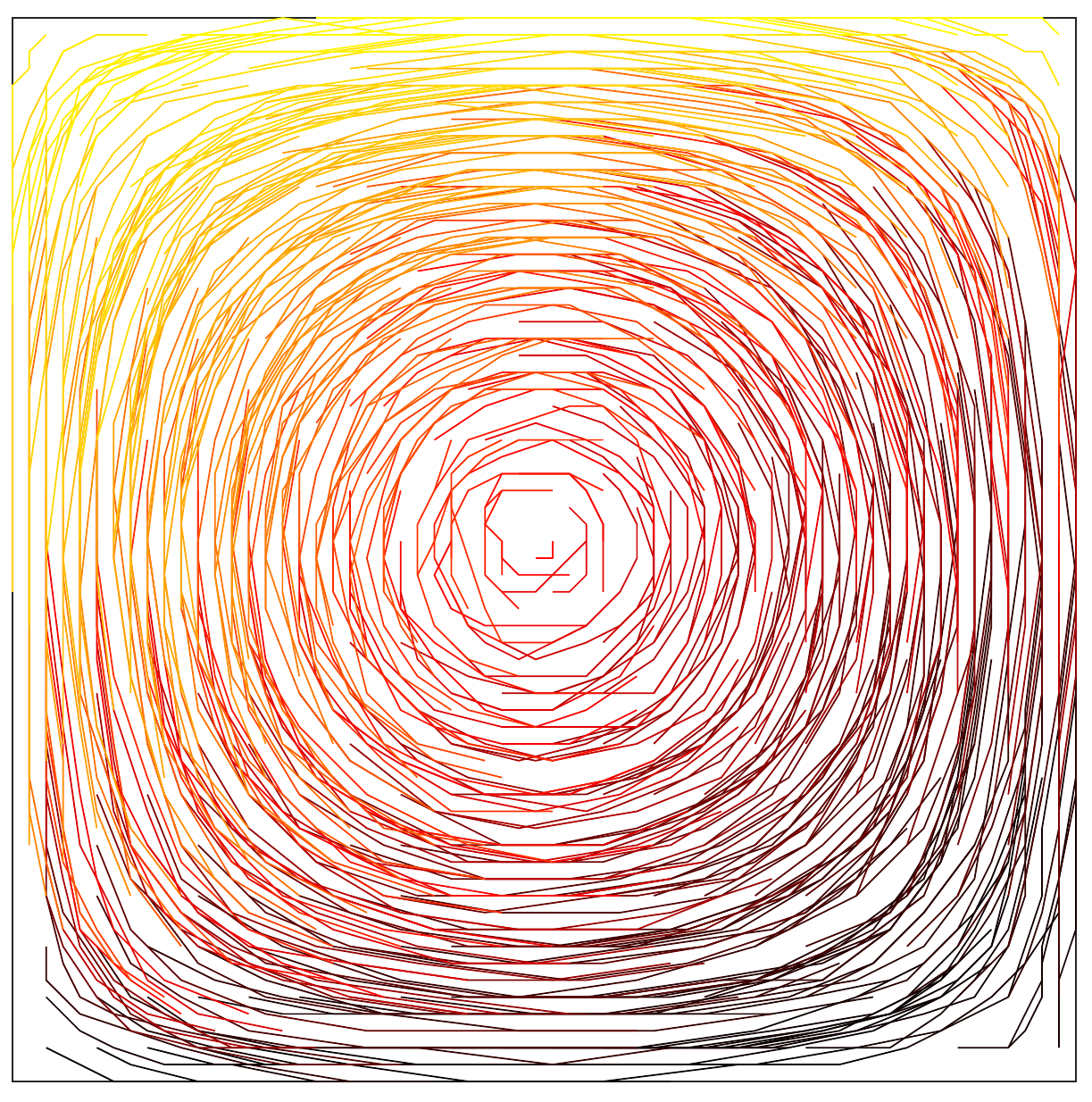}\\
    \centerline{$\tf<t_\star$}
    \\[10pt]
    \includegraphics[width=\textwidth]{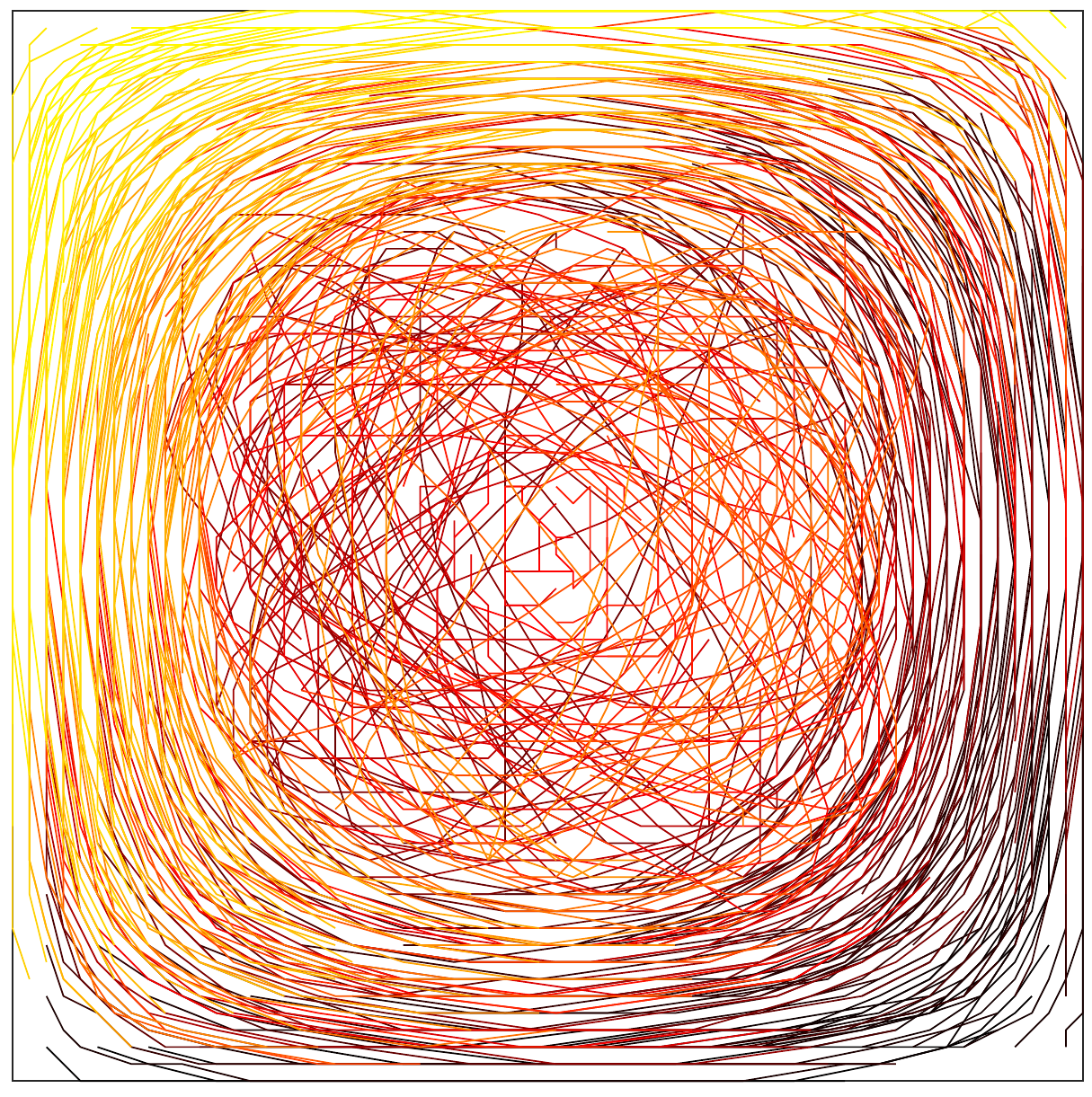}\\
    \centerline{$\tf>t_\star$}
  \end{minipage}
  \end{center}
  \caption{Left: Average Lagrangian displacement for both $\tf<t_\star$ (green) $\tf>t_\star$ (red). Right: realisations of the Lagrangian trajectories, again for time lags $\tf$ below (top) and above (down) the critical time $t_\star$. \add{The colorscale there codes the initial vertical positions of the trajectory.}}
  \label{fig:beltramilagrangian}
\end{figure}

Perhaps, this is even better seen from the Lagrangian measurements shown in Fig.~\ref{fig:beltramilagrangian}. As can be seen on the left-hand panel, the motions of fluid particles transit from a diffusive towards a faster ballistic regime when $\tf$ is increased. The present breakdown of the Lagrangian flow is therefore not related to the phenomenon of explosive turbulent separation, nor to Richardson's diffusion. The origin of a ballistic behaviour can actually be identified by looking at specific realisations of the generalised Beltrami trajectories.  One indeed observes on the right-hand panels of Fig.~\ref{fig:beltramilagrangian} that for $\tf > t_\star$ that trajectories located at the center of the domain ``tunnel'' through the large-scale vortex  to reach almost ballistically their final prescribed location. The statistics is then dominated by such displacements, which prove more efficient at transporting energy than classical rotating motions.  Deviations to a purely ballistic regime come from the Lagrangian trajectories in a thin layer near the domain boundary, which remain ``classical'': This is likely a consequence of the non-uniform distribution in Beltrami flows of vorticity, which is weaker at the boundary than at the center of the domain. 

\subsection{Comments}
The statistical analysis on generalised Beltrami flows confirms the intuition built from the $(2\times2)$ solid-rotation model considered in \S\ref{sec:Fourcell}: The transition from deterministic to non-deterministic generalised flows is not at any rate a transition to turbulence, however appealing this would have sounded. The generalised flows above $t_\star$ display a number of non-physical features, ranging from non-monotonous sharp variations of the energy at initial and final times, to ``vortex-tunneling'' ballistic motions of fluid trajectories. These highly undesirable specificities could possibly be avoided by further constraining the Lagrangian variational principle.  For example, one can break the time-symmetric formulation by enforcing that energy be decreasing. On the one hand, this path could likely be a dead-end as it is clear, already for Arnold's regular geodesic, that the ``$t_\star$ barrier'' is a defect of the boundary-value formulation, but not of the least-action principle. On the other hand, further constraining the Lagrangian trajectories may allow to formulate Brenier's principle as a non-trivial free-end optimisation problem (namely an initial value problem in the language of partial differential equations), hereby wiping off artefacts due to the boundary-value formulation. At this point, our results therefore suggest that variational reconstructions of intermediate (turbulent) dynamics cannot rely solely on (\ref{eq:BVP}). It also requires an amount of data that ensures sufficiently close time-boundary conditions. Only at this price can one obtain a non-spurious ``$t<t_\star$'' behaviour. With that in mind, we next apply least-action principles to turbulent data.

\section{Generalised flows and decaying 2D turbulence}
\label{sec:decay}

The next step of our approach is to consider even more realistic cases where vorticity varies not only in space, but also in time.  This is natural when one has in mind turbulent dynamics. Yet, the least-action principle that we have discussed is restricted to velocity fields solving the Euler equations. The classical solutions that were considered so far are of limited relevance to turbulent flows. In particular, they are too regular to occur as non-trivial limits of solutions to the Navier--Stokes equations when the viscosity $\nu\to0$.  In that context, it is believed that weak (irregular) solutions to the Euler equations are more suitable candidates.  Generalised least-action principle would be particularly useful if they could be used to tackle such non-smooth velocity fields.  We naturally have in mind the possibility of addressing three-dimensional flows, but we here restrict ourselves to the problem of the two-dimensional direct enstrophy cascade.

It has been known since \cite{batchelor1969computation} that two-dimensional turbulence does not dissipate kinetic energy when the viscosity tends to zero. In the absence of forcing, there is however another inviscid invariant, namely the enstrophy $Z$ defined as the space integral of the square of the vorticity $\omega = \partial_xv_y-\partial_yv_x$.  Batchelor's scaling predicts that the enstrophy displays a dissipative anomaly but this requires approaching the inviscid limit with an infinite enstrophy \citep[see][]{eyink2001dissipation}.  While these questions are still not fully settled \citep[see][]{lopes2006weak,dmitruk2005numerical,vigdorovich2018enstrophy}, one can reasonably expect that two-dimensional, decaying, turbulent velocity fields are irregular solutions to the Euler equations. This, together with the absence of kinetic energy dissipation, motivates applying to that case the generalised least-action principle. Let us note that the presence of a finite enstrophy dissipation in the limit of vanishing viscosity makes the dynamics time irreversible. It is a priori unclear if and how this feature will be accommodated by Brenier's principle.

\begin{figure}
  \centerline{\includegraphics[width=1.2\textwidth]{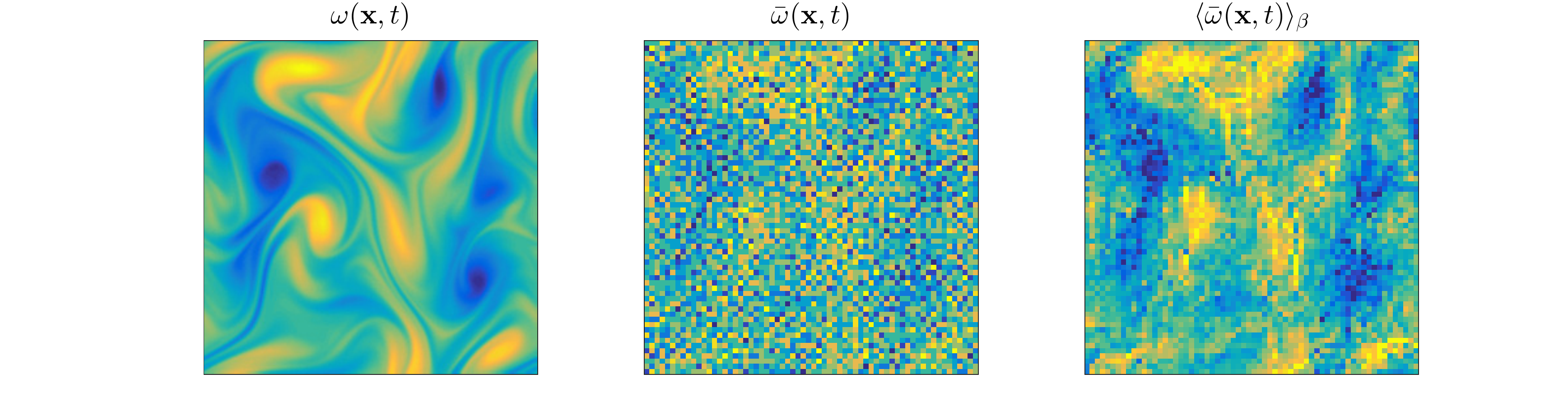}}
  \caption{\label{fig:three_vorticitie} Snapshot  of the vorticity field at time $t=\tf/2$. Positive values are shown in blue and negative ones in yellow. The left-hand panel is the original vorticity field obtained from the direct numerical simulation with $256^2$ collocation points. The centre panel is a fluctuating realisation of the reconstructed vorticity on a $64^2$ coarse graining with a finite temperature $\beta^{-1} = 1$. The right-hand panel is the generalised flow approximation obtained by averaging over the Gibbs ensemble. }
\end{figure}
We perform direct numerical simulations of the decaying two-dimensional Navier--Stokes equations. The initial velocity field, with energy $E(0)$ and enstrophy $Z(0)$, is chosen from another set of simulations where both a large-scale random forcing and linear friction are present. We use a pseudo-spectral solver with $256^2$ collocation points and kinematic viscosity $\nu = 10^{-3}$. We embed $250,000$ tracers and compute permutation flows $\{ \bsigma_n\}$ on a $N_x^2 = 64^2$ coarse-grained grid every $\tf = Z^{-1/2}(0)$. This is the smaller timescale of the flow. The two-dimensional direct cascade is indeed characterised by a single turnover time prescribed by the enstrophy.

The obtained permutation flow provides boundary conditions for the generalised least-action principle.  We use the Metropolis algorithm of \S\ref{subsec:permutations} to reconstruct the velocity field at five intermediate times between $n\, \tf$ and $(n+1)\, \tf$. Figure~\ref{fig:three_vorticitie} illustrates this procedure for $n=0$.  The vorticity field at time $\tf$ obtained from the direct simulations is shown on the left-hand panel. The central panel shows a realisation of the reconstructed vorticity $\bar{\omega}$ for an inverse temperature $\beta=1$, and without any averaging. It is obtained by finite differences of the Lagrangian displacement, and one can hardly distinguish any structure. It is only after averaging over $10^6$ Monte-Carlo times that one smears out the noise.  The spatial organisation of the original flow is indeed recovered on the right-hand panel of Fig.~\ref{fig:three_vorticitie}, which displays the Gibbs-averaged vorticity $\langle\bar{\omega}\rangle_\beta$. The blurry aspect is due to the finite value of temperature. The generalised vorticity will be formally obtained for $N_x,N_t\to\infty$, together with $\beta \to\infty$.

\begin{figure}
  \centerline{\includegraphics[width=0.49\textwidth]{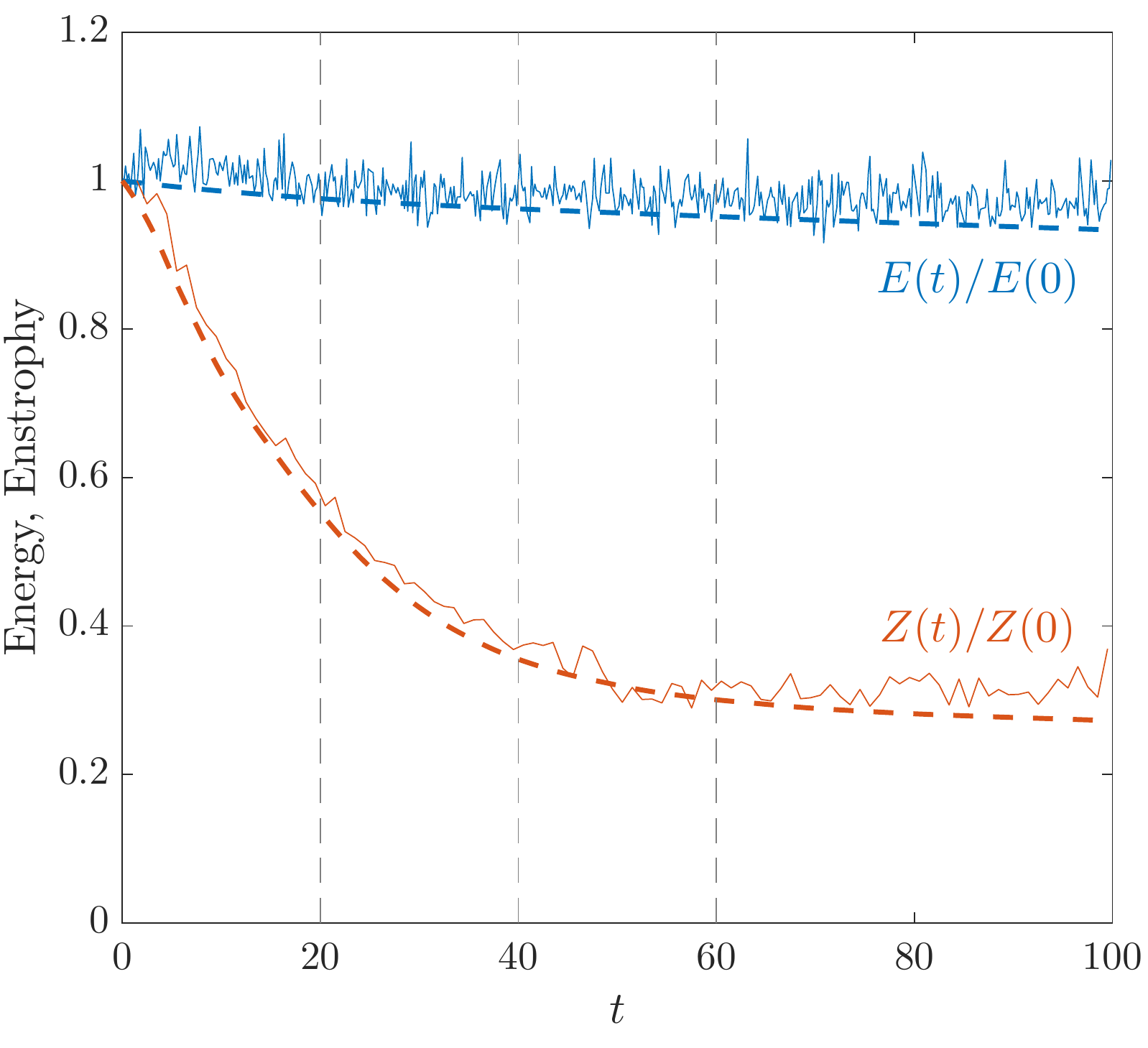}
    \includegraphics[width=0.49\textwidth]{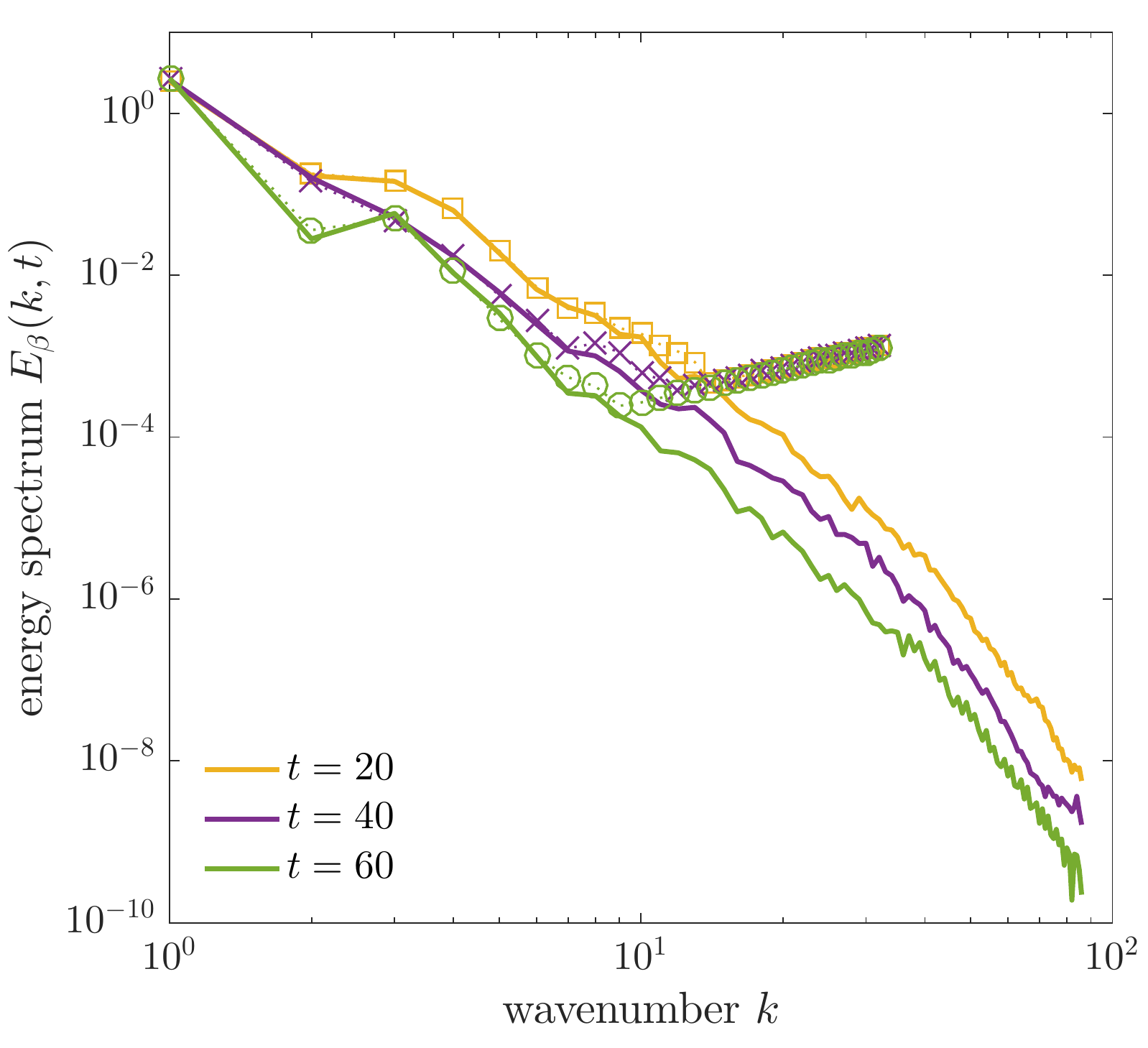}}
  \caption{\label{fig:decay} Left: Energy and enstrophy as a function of time both for the
    actual field (dashed lines) and the reconstructed one (thin solid lines). Here, $Z(t)$ is averaged over
    the reconstructed time intervals $[n\,\tf,(n+1)\,\tf]$, while $E(t)$ is not, whence the stronger
    fluctuations for the second. Time is here in units of $Z^{-1/2}(0)$. Right: kinetic energy spectra of both the original velocity field (solid lines) and of the reconstructed one (dashed lines with symbols) at different times, as labeled.}
\end{figure}
This procedure is repeated up to $n=100$. Figure~\ref{fig:decay} (left) shows the time evolution of the kinetic energy $E(t)$ and of the enstrophy $Z(t)$. Data from the direct simulation (shown as dashed lines) show that the enstrophy decreases abruptly while energy is almost conserved. This confirms that our settings correspond to the small-viscosity limit of two-dimensional decaying turbulence.  The slow decrease of $E(t)$ is due to the tiny amount of viscous dissipation used in the numerics.  The solid lines are the energy and enstrophy obtained from the reconstructed flow. We observe that the global decay is properly reproduced with superimposed fluctuations due to a finite temperature. Other deviations might be due to statistical errors in the ensemble average or to the discreteness of the reconstructed field. The scaling properties of the velocity fields can be examined from the energy power spectra shown in the right-hand panel of Fig.~\ref{fig:decay}. Data show an excellent agreement at large scales. As observed earlier in Beltrami flows, reconstructed fields display a $\propto k$ range at small scales. The crossover wavenumber apparently decreases with time, so that one can expect a full thermalisation at large times. The quality of the reconstruction will thus deteriorate, as can be already inferred from the left-hand panel at times $t\gtrsim70$.

These attempts at using Brenier's principle to reconstruct weak solutions to the Euler equations are rather promising. In particular, this approach seems suited to capture irreversible dynamics, and is hence compatible with the concept of dissipative inviscid solutions. Nevertheless, one should be aware that the reconstructed data is here very regular compared to a genuine turbulent three-dimensional flow. While this still constitutes a first step, the observables that we considered are dominated by the larger scales, with a velocity field that apparently remains very close to a classical solution of the inviscid dynamics.

\section{Concluding remarks}
\label{sec:conclusion}
\add{From a general viewpoint, our analysis highlights one of the limitations of least-action principles when formulated as  two-end  boundary-value problems in time: They fail to reconstruct vortical dynamics over long time lags. In the solid rotation example, the issue relates to the final map being likely degenerate, in the sense that can be concomitantly compatible with different equally energetic strong Euler solutions.
The critical time $t_\star$ over which this phenomenon happen  is naturally interpreted as a \emph{turnover time}, and is a particular case  of Brenier's general formula  (\ref{eq:breniertime}).}

\add{Because Brenier's generalised framework is intrinsically probabilistic, it guarantees the existence of unique minimisers for all times. Yet, the non-deterministic properties of the minimising flows that appear when $\tf \ge t_\star$ are essentially an artefact of the boundary-value formulation itself. This finding appears unambiguously through the Beltrami example considered in \S\ref{sec:Beltrami}.  In the regime $\tf \ge t_\star$, the statistical features of the minimising generalised flows are found to be unphysical and therefore unrelated to any kind of turbulent measure.  To be used faithfully, it is therefore necessary to restrict the use of the generalised least action principle, and only solve it for short time lags, caracterised by $\tf<t_\star$.  }

\add{
In the regime $\tf<t_\star$, our work  illustrates that Brenier's generalised least-action principle (\ref{eq:BVP}) is then not only useful but also versatile.
Specifically, our numerics show that it can be efficiently  invoke to  coarse-grain  various types of inviscid dynamics in terms of generalised flows. Coarse-grained statistics for both the Lagrangian trajectories and for the full velocity field can then be reconstructed.} 
\add{Essentially, the generalised framework is practical  because it considerably relaxes Arnold's}. Compared to Arnold's formulation (\ref{eq:AVP}), the optimisation problem (\ref{eq:BVP})  is indeed largely unconstrained,  and in particular, the probabilistic framework does not \add{rely on}  any assumption about the smoothness of the minimising flow. This \add{highly} desirable feature makes it particularly suitable  as a coarse-graining tool,  as Brenier's formulation naturally carries through on discrete settings, both in space and in time\,---\,at sharp contrast with Arnold's formulation \add{ based on time-differentiable diffeomorphisms}. Besides, the optimising space is sufficiently large, so that minimising algorithms can converge efficiently, without getting trapped within local minima.
\add{In the permutation flow viewpoint that we adopted, we note that the number of microscopic configurations to be optimised upon is approximately $10^{1025}$, to yield a space-time coarse-graining with $64^2$ gridpoints and 7 timesteps.}
\add{We see this as the main reason why  simple Monte-Carlo strategies, such as the one presented in \S\ref{sec:statphys}, yield faithful approximates to (\ref{eq:BVP}).}
 
The inviscid dynamics that lie within the scope of Brenier's action principle include the strong solutions captured by Arnold's principle, in which case the minimising generalised flow becomes degenerate as it concentrates on the deterministic solution.
More interestingly, the numerical analysis presented in \S\ref{sec:decay} suggests that Brenier's principle may also reproduce irreversible dynamics that are a priori prescribed by weak solutions to the Euler equation.  
\add{While the case of 2D decaying turbulence is peculiar,  in the sense that the dynamics is smooth and dominated by large-scale eddies,
this numerical evidence is in our view  very  promising. In particular, it paves the way towards new tools where generalised flows are used to coarse-grain truly multi-scale hydrodynamics, including fully developed turbulence. }

This overall perspective naturally  relies on the crucial  assumption that turbulence is described by weak solutions to the Euler equations. If such was indeed the case, generalised flows would then provide a natural framework to think of turbulent solutions in terms of Lagrangian transition probabilities.

\bibliographystyle{jfm} 
\bibliography{biblio}

\begin{thebibliography}{54}
\expandafter\ifx\csname natexlab\endcsname\relax\def\natexlab#1{#1}\fi

\bibitem[Abarbanel {\em et~al.\/}(1986)Abarbanel, Holm, Marsden \&
  Ratiu]{abarbanel1986nonlinear}
{\sc Abarbanel, H.~D., Holm, D.~D., Marsden, J.~E. \& Ratiu, T.~S.} 1986
  Nonlinear stability analysis of stratified fluid equilibria. {\em Phil.
  Trans. R. Soc. Lond. A\/} {\bf 318}~(1543), 349--409.

\bibitem[Antonia \& Burattini(2006)]{antonia2006approach}
{\sc Antonia, R.~A. \& Burattini, P.} 2006 Approach to the 4/5 law in
  homogeneous isotropic turbulence. {\em J. Fluid Mech.\/} {\bf 550}, 175--184.

\bibitem[Arnold(1966)]{arnold1966geometrie}
{\sc Arnold, V.~I.} 1966 Sur la g{\'e}om{\'e}trie diff{\'e}rentielle des
  groupes de lie de dimension infinie et ses applicationsa l’hydrodynamique
  des fluides parfaits. {\em Ann. Inst. Fourier\/} {\bf 16}, 319--361.

\bibitem[Arnold(2013)]{arnold2013mathematical}
{\sc Arnold, V.~I.} 2013 {\em Mathematical methods of classical mechanics\/},
  {\em Graduate Texts in Mathematics\/}, vol.~60. New York: Springer.

\bibitem[Batchelor(1969)]{batchelor1969computation}
{\sc Batchelor, G.~K.} 1969 Computation of the energy spectrum in homogeneous
  two-dimensional turbulence. {\em Phys. Fluids\/} {\bf 12}, II--233.

\bibitem[Baxter(2016)]{baxter2016exactly}
{\sc Baxter, Rodney~J} 2016 {\em Exactly solved models in statistical
  mechanics\/}. Elsevier.

\bibitem[Benamou {\em et~al.\/}(2015)Benamou, Carlier, Cuturi, Nenna \&
  Peyr{\'e}]{benamou2015iterative}
{\sc Benamou, J.-D., Carlier, G., Cuturi, M., Nenna, L. \& Peyr{\'e}, G.} 2015
  Iterative {B}regman projections for regularized transportation problems. {\em
  SIAM J. Sci. Comput.\/} {\bf 37}, A1111--A1138.

\bibitem[Benamou {\em et~al.\/}(2017)Benamou, Carlier \&
  Nenna]{benamou2017generalized}
{\sc Benamou, J.-D., Carlier, G. \& Nenna, L.} 2017 Generalized incompressible
  flows, multi-marginal transport and {S}inkhorn algorithm. ArXiv preprint
  arXiv:1710.08234.

\bibitem[Binder(1986)]{binder1986introduction}
{\sc Binder, K.} 1986 Introduction: {T}heory and “technical” aspects of
  {M}onte {C}arlo simulations. In {\em Monte Carlo Methods in Statistical
  Physics\/}, pp. 1--45. Berlin, Heidelberg: Springer.

\bibitem[Bouchet \& Corvellec(2010)]{bouchet2010invariant}
{\sc Bouchet, F. \& Corvellec, M.} 2010 Invariant measures of the 2d euler and
  vlasov equations. {\em J. Stat. Mech. Theory Exp.\/} {\bf 2010}, P08021.

\bibitem[Bouchet \& Venaille(2012)]{bouchet2012statistical}
{\sc Bouchet, F. \& Venaille, A.} 2012 Statistical mechanics of two-dimensional
  and geophysical flows. {\em Phys. Rep.\/} {\bf 515}, 227--295.

\bibitem[Brenier(1989)]{brenier1989least}
{\sc Brenier, Y.} 1989 The least action principle and the related concept of
  generalized flows for incompressible perfect fluids. {\em J. Am. Math.
  Soc.\/} {\bf 2}, 225--255.

\bibitem[Brenier(1999)]{brenier1999minimal}
{\sc Brenier, Y.} 1999 Minimal geodesics on groups of volume-preserving maps
  and generalized solutions of the {E}uler equations. {\em Commun. Pure Appl.
  Math.\/} {\bf 52}, 411--452.

\bibitem[Brenier(2008)]{brenier2008generalized}
{\sc Brenier, Y.} 2008 Generalized solutions and hydrostatic approximation of
  the {E}uler equations. {\em Physica D\/} {\bf 237}, 1982--1988.

\bibitem[Brenier {\em et~al.\/}(2011)Brenier, De~Lellis \&
  Sz{\'e}kelyhidi]{brenier2011weak}
{\sc Brenier, Y., De~Lellis, C. \& Sz{\'e}kelyhidi, L.} 2011 Weak-strong
  uniqueness for measure-valued solutions. {\em Commun. Math. Phys\/} {\bf
  305}, 351--361.

\bibitem[Buckmaster(2015)]{buckmaster2015onsager}
{\sc Buckmaster, T.} 2015 Onsager’s conjecture almost everywhere in time.
  {\em Commun. Math. Phys\/} {\bf 333}, 1175--1198.

\bibitem[Buckmaster {\em et~al.\/}(2016)Buckmaster, Lellis \&
  Sz{\'e}kelyhidi]{buckmaster2016dissipative}
{\sc Buckmaster, T., Lellis, C. \& Sz{\'e}kelyhidi, L.} 2016 Dissipative
  {E}uler flows with {O}nsager-critical spatial regularity. {\em Commun. Pure
  Appl. Math\/} {\bf 69}, 1613--1670.

\bibitem[DiPerna \& Majda(1987)]{diperna1987oscillations}
{\sc DiPerna, R.~J. \& Majda, A.~J.} 1987 Oscillations and concentrations in
  weak solutions of the incompressible fluid equations. {\em Commun. Math.
  Phys\/} {\bf 108}, 667--689.

\bibitem[Dmitruk \& Montgomery(2005)]{dmitruk2005numerical}
{\sc Dmitruk, P. \& Montgomery, D.~C.} 2005 Numerical study of the decay of
  enstrophy in a two-dimensional navier--stokes fluid in the limit of very
  small viscosities. {\em Phys. Fluids\/} {\bf 17}, 035114.

\bibitem[Duchon \& Robert(2000)]{duchon2000inertial}
{\sc Duchon, J. \& Robert, R.} 2000 Inertial energy dissipation for weak
  solutions of incompressible {E}uler and {N}avier-{S}tokes equations. {\em
  Nonlinearity\/} {\bf 13}, 249--255.

\bibitem[Ebin \& Marsden(1970)]{ebin1970groups}
{\sc Ebin, D.~G. \& Marsden, J.} 1970 Groups of diffeomorphisms and the motion
  of an incompressible fluid. {\em Ann. Math.\/} {\bf 92}, 102--163.

\bibitem[Eyink(2001)]{eyink2001dissipation}
{\sc Eyink, G.~L.} 2001 Dissipation in turbulent solutions of 2{D} {E}uler
  equations. {\em Nonlinearity\/} {\bf 14}, 787--802.

\bibitem[Eyink(2002)]{eyink2002local}
{\sc Eyink, G.~L.} 2002 Local 4/5-law and energy dissipation anomaly in
  turbulence. {\em Nonlinearity\/} {\bf 16}, 137--145.

\bibitem[Eyink \& Sreenivasan(2006)]{eyink2006onsager}
{\sc Eyink, G.~L. \& Sreenivasan, K.~R.} 2006 Onsager and the theory of
  hydrodynamic turbulence. {\em Rev. Mod. Phys.\/} {\bf 78}, 87--135.

\bibitem[Falkovich {\em et~al.\/}(2001)Falkovich, Gaw\c{e}dzki \&
  Vergassola]{falkovich2001particles}
{\sc Falkovich, G., Gaw\c{e}dzki, K. \& Vergassola, M.} 2001 Particles and
  fields in fluid turbulence. {\em Rev. Mod. Phys.\/} {\bf 73}, 913--975.

\bibitem[Farazmand \& Serra(2018)]{farazmand2018variational}
{\sc Farazmand, Mohammad \& Serra, Mattia} 2018 Variational {L}agrangian
  formulation of the {E}uler equations for incompressible flow: A simple
  derivation. {\em arXiv preprint arXiv:1807.02726\/} .

\bibitem[Gallou{\"e}t \& M{\'e}rigot(2018)]{gallouet2016lagrangian}
{\sc Gallou{\"e}t, T. \& M{\'e}rigot, Q.} 2018 A lagrangian scheme \`{a} la
  {B}renier for the incompressible {E}uler equations. {\em Foundations of
  Computational Mathematics\/} {\bf 18}, 835--865.

\bibitem[Gaw\c{e}dzki(2001)]{gawedzki2001turbulent}
{\sc Gaw\c{e}dzki, K.} 2001 Turbulent advection and breakdown of the lagrangian
  flow. In {\em Intermittency in Turbulent Flows\/} (ed. J.C. Vassilicos), pp.
  86--104. Cambridge: Cambridge University Press.

\bibitem[Holm \& Kupershmidt(1983)]{holm1983noncanonical}
{\sc Holm, D.~D. \& Kupershmidt, B.~A.} 1983 Noncanonical {H}amiltonian
  formulation of ideal magnetohydrodynamics. {\em Physica D\/} {\bf 7},
  330--333.

\bibitem[Holm {\em et~al.\/}(1985)Holm, Marsden, Ratiu \&
  Weinstein]{holm1985nonlinear}
{\sc Holm, D.~D., Marsden, J.~E., Ratiu, T. \& Weinstein, A.} 1985 Nonlinear
  stability of fluid and plasma equilibria. {\em Phys. Rep.\/} {\bf 123},
  1--116.

\bibitem[Isett(2018)]{isett2016proof}
{\sc Isett, P.} 2018 A proof of {O}nsager's conjecture. {\em Ann. Math.\/} {\bf
  188}, 871--963.

\bibitem[Jos{\'e} \& Saletan(2000)]{jose2000classical}
{\sc Jos{\'e}, J. \& Saletan, E.} 2000 {\em Classical dynamics: a contemporary
  approach\/}. Cambridge: Cambridge University Press.

\bibitem[Khesin \& Arnold(2005)]{khesin2005topological}
{\sc Khesin, B. \& Arnold, V.~I.} 2005 Topological fluid dynamics. {\em Notices
  Am. Math. Soc.\/} {\bf 52}, 9--19.

\bibitem[Kraus {\em et~al.\/}(2016)Kraus, Tassi \&
  Grasso]{kraus2016variational}
{\sc Kraus, M., Tassi, E. \& Grasso, D.} 2016 Variational integrators for
  reduced magnetohydrodynamics. {\em J. Comput. Phys.\/} {\bf 321}, 435--458.

\bibitem[Lopes~Filho {\em et~al.\/}(2006)Lopes~Filho, Mazzucato \&
  Nussenzveig~Lopes]{lopes2006weak}
{\sc Lopes~Filho, M.~C., Mazzucato, A.~L. \& Nussenzveig~Lopes, H.~J.} 2006
  Weak solutions, renormalized solutions and enstrophy defects in 2d
  turbulence. {\em Arch. Ration. Mech. Anal.\/} {\bf 179}, 353--387.

\bibitem[Marsden \& West(2001)]{marsden2001discrete}
{\sc Marsden, J.~E. \& West, M.} 2001 Discrete mechanics and variational
  integrators. {\em Acta Numer.\/} {\bf 10}, 357--514.

\bibitem[M{\'e}rigot \& Mirebeau(2016)]{merigot2016minimal}
{\sc M{\'e}rigot, Q. \& Mirebeau, J.-M.} 2016 Minimal geodesics along
  volume-preserving maps, through semidiscrete optimal transport. {\em SIAM J.
  Numer. Anal.\/} {\bf 54}, 3465--3492.

\bibitem[Miller {\em et~al.\/}(1992)Miller, Weichman \&
  Cross]{miller1992statistical}
{\sc Miller, J., Weichman, P.~B. \& Cross, M.~C.} 1992 Statistical mechanics,
  {E}uler’s equation, and {J}upiter’s red spot. {\em Phys. Rev. A\/} {\bf
  45}, 2328--2359.

\bibitem[Morrison(1998)]{morrison1998hamiltonian}
{\sc Morrison, P.~J.} 1998 Hamiltonian description of the ideal fluid. {\em
  Rev. Mod. Phys.\/} {\bf 70}, 467--521.

\bibitem[Morrison(2005)]{morrison2005hamiltonian}
{\sc Morrison, P.~J.} 2005 Hamiltonian and action principle formulations of
  plasma physics. {\em Phys. Plasmas\/} {\bf 12}, 058102.

\bibitem[Nenna(2016)]{nenna2016numerical}
{\sc Nenna, L.} 2016 Numerical methods for multi-marginal optimal
  transportation. PhD thesis, PSL Research University.

\bibitem[Nocedal \& Wright(2006)]{nocedal2006numerical}
{\sc Nocedal, Jorge \& Wright, Stephen} 2006 {\em Numerical optimization\/}.
  Springer Science \& Business Media.

\bibitem[Onsager(1949)]{onsager1949statistical}
{\sc Onsager, L.} 1949 Statistical hydrodynamics. {\em Nuovo Cimento\/} {\bf
  6}, 279--287.

\bibitem[Robert \& Sommeria(1991)]{robert1991statistical}
{\sc Robert, R. \& Sommeria, J.} 1991 Statistical equilibrium states for
  two-dimensional flows. {\em J. Fluid Mech.\/} {\bf 229}, 291--310.

\bibitem[Salmon(1983)]{salmon1983practical}
{\sc Salmon, R.} 1983 Practical use of {H}amilton's principle. {\em J. Fluid
  Mech.\/} {\bf 132}, 431--444.

\bibitem[Salmon(1985)]{salmon1985new}
{\sc Salmon, R.} 1985 New equations for nearly geostrophic flow. {\em J. Fluid
  Mech.\/} {\bf 153}, 461--477.

\bibitem[Salmon(1988)]{salmon1988hamiltonian}
{\sc Salmon, R.} 1988 Hamiltonian fluid mechanics. {\em Annual Rev. Fluid
  Mech.\/} {\bf 20}, 225--256.

\bibitem[Saw {\em et~al.\/}(2018)Saw, Debue, Kuzzay, Daviaud \&
  Dubrulle]{saw2018universality}
{\sc Saw, E.-W., Debue, P., Kuzzay, D., Daviaud, F. \& Dubrulle, B.} 2018 On
  the universality of anomalous scaling exponents of structure functions in
  turbulent flows. {\em J. Fluid Mech.\/} {\bf 837}, 657--669.

\bibitem[Shepherd(1990)]{shepherd1990symmetries}
{\sc Shepherd, T.~G.} 1990 Symmetries, conservation laws, and {H}amiltonian
  structure in geophysical fluid dynamics. In {\em Adv. Geophys.\/}, , vol.~32,
  pp. 287--338. Elsevier.

\bibitem[Shnirelman(2000)]{shnirelman2000weak}
{\sc Shnirelman, A.} 2000 Weak solutions with decreasing energy of
  incompressible euler equations. {\em Commun. Math. Phys.\/} {\bf 210},
  541--603.

\bibitem[Sreenivasan(1984)]{sreenivasan1984scaling}
{\sc Sreenivasan, K.~R.} 1984 On the scaling of the turbulence energy
  dissipation rate. {\em Physics FRluids\/} {\bf 27}, 1048--1051.

\bibitem[Vigdorovich(2018)]{vigdorovich2018enstrophy}
{\sc Vigdorovich, I.} 2018 Enstrophy spectrum in freely decaying
  two-dimensional self-similar turbulent flow. {\em Phys. Rev. E\/} {\bf 98},
  033110.

\bibitem[Villani(2009)]{villani2008optimal}
{\sc Villani, C.} 2009 {\em Optimal transport: old and new\/}, {\em Grundlehren
  der mathematischen Wissenschaften\/}, vol. 338. Berlin Heidelberg:
  Springer-Verlag.

\bibitem[Zeitlin(2004)]{zeitlin2004self}
{\sc Zeitlin, V.} 2004 Self-consistent finite-mode approximations for the
  hydrodynamics of an incompressible fluid on nonrotating and rotating spheres.
  {\em Phys. Rev. Lett.\/} {\bf 93}, 264501.

\end{thebibliography}
\end{document}